\documentclass{aa}  
\usepackage{graphicx}
\usepackage{subcaption}
\usepackage{placeins} 
\usepackage{float}
\usepackage{tabularx}
\usepackage[table]{xcolor}
\usepackage{txfonts}
\usepackage{lscape}
\usepackage{booktabs}
\usepackage{threeparttablex}
\usepackage{wasysym}
\usepackage{arydshln}
\usepackage{mathrsfs}
\usepackage{amssymb}
\usepackage{color}
\usepackage[dvipsnames]{xcolor}
\usepackage[normalem]{ulem}
\usepackage[bookmarks=false,colorlinks=true,linkcolor=black,citecolor=cyan,filecolor=black,urlcolor=cyan]{hyperref}
\usepackage{stfloats}

\begin{document}

\title{Data span and frequency coverage requirements for robust detection and inference in PTAs}
\subtitle{A case study with EPTA \texttt{DR2}}

\author{
Irene Ferranti,$^{1,2}$\thanks{E-mail: i.ferranti@campus.unimib.it}
Mikel Falxa,$^{1}$
Federico Fantoccoli,$^{5}$
Alberto Sesana,$^{1,2,3}$
Golam Shaifullah,$^{1,2,4}$}

\date{Accepted XXX. Received YYY; in original form ZZZ}
\institute{
    Dipartimento di Fisica ``G. Occhialini", Università degli Studi di Milano-Bicocca, Piazza della Scienza 3, I-20126 Milano, Italy
    \and
    INFN, Sezione di Milano-Bicocca, Piazza della Scienza 3, I-20126 Milano, Italy
    \and
    INAF - Osservatorio Astronomico di Brera, via Brera 20, I-20121 Milano, Italy
    \and
    INAF - Osservatorio Astronomico di Cagliari, via della Scienza 5, 09047 Selargius (CA), Italy
    \and
    Max Planck Institute for Gravitational Physics (Albert Einstein Institute), Am Mühlenberg 1, D-14476 Potsdam, Germany
    }

\titlerunning{Data span and frequency coverage requirements for robust detection and inference in PTAs}
\authorrunning{Ferranti et al.}

\abstract{Pulsar Timing Arrays (PTAs) are approaching the sensitivity required to make a 5-$\sigma$ detection of the nanohertz stochastic gravitational-wave background (GWB), making it crucial to develop a comprehensive understanding of our data and of the outcomes of our analysis pipelines. Thus, it becomes essential to understand a counterintuitive feature revealed in the recent results from the European Pulsar Timing Array (EPTA) second data release (\texttt{DR2}): when restricting the dataset to its ultimate $\sim$10.3 years (\texttt{DR2new}), the inferred GWB significance increases from $\lesssim2\sigma$ for the full 25-year dataset (\texttt{DR2full}), to $\gtrsim3.5\sigma$ for \texttt{DR2new}. In this work, we investigate whether this behaviour reflects an anomaly in the data or it is a possible outcome of the analysis pipeline. Using realistic \texttt{DR2}-like simulations, we generate multiple realisations with varying observation timespans and analyse their impact on GWB detection and parameter estimation. We find that the first $\sim$10 years of \texttt{DR2} contribute little to the GWB evidence due to the limited bandwidth of the observation frequency, leading to a significant overlap between the \texttt{DR2full} and \texttt{DR2new} HD S/N distributions. As a consequence, random noise fluctuations result in \texttt{DR2new} producing a higher GWB significance than \texttt{DR2full} in $\sim15\%$ of the cases and 5$\%$ of the realisations are compatible with the HD S/N of the real \texttt{DR2full} and \texttt{DR2new}.
This suggest that what is observed in the data, although unlikely, is consistent with being a $\sim 2\sigma$ outcome due to noise fluctuations.
We also find that, regardless of the significance, \texttt{DR2new} simulated data yield biased GWB parameter estimates, primarily due to spectral leakage effects that are disregarded in the analysis and tend to flatten the inferred power-law spectrum. Including leakage in the model, returns unbiased parameter estimates, demonstrating that \texttt{DR2new} is reliable when the signal is appropriately modeled. Furthermore, we show that combining EPTA \texttt{DR2full} data with complementary long-baseline observations from NANOGrav and PPTA and with low frequency observations from LOFAR and NenuFAR significantly improves the GWB evidence and precision and accuracy in the parameter estimation, supporting the case of combining \texttt{DR2full} within the IPTA framework. Finally, we explore more in detail the impact of the observation timespan on parameter estimation, focusing on long-baseline (25yr) and short-baseline (5yr) datasets. We find that short-baseline datasets tend to introduce significant bias towards high amplitudes in the estimation of GWB parameters, and the short time span makes them very ineffective at constraining the GWB slope.}
    

\maketitle


\section{Introduction}

At nanohertz (nHz) frequencies, a stochastic gravitational-wave background (GWB) is expected to arise from a cosmic population of supermassive black hole binaries (SMBHBs; \citealt{1995ApJ...446..543R,Jaffe2003,2003ApJ...590..691W,2008MNRAS.390..192S}), as well as from a variety of processes occurring in the early-Universe \citep[see][and references therein]{2018CQGra..35p3001C,2023ApJ...951L..11A,2024A&A...685A..94E}. By conducting high-precision timing of millisecond pulsars, pulsar timing arrays (PTAs; \citealt{1990ApJ...361..300F}) produce datasets sensitive enough to detect tiny perturbations in spacetime curvature caused by passing GWs, making them sensitive to potential nHz GWBs. Recently, the European PTA (EPTA), in partnership with the Indian PTA (InPTA) \citep{wm3}, along with the North American Nanohertz Observatory for Gravitational Waves (NANOGrav; \citealt{NANOGrav}), the Parkes PTA (PPTA; \citealt{PPTA}), the Chinese PTA (CPTA; \citealt{CPTA}) and the MeerKAT Pulsar Timing Array (MPTA; \citealt{2025MNRAS.536.1489M}) reported evidence of a GWB-like signal with a significance of 1–4$ \sigma$ (depending on the dataset), effectively opening the first observational window onto the nHz GW sky.

A PTA dataset consists of timing residuals, $\delta\vec{t}$, obtained by subtracting a best-fit timing model (TM) from the observed pulses' times of arrival (TOAs) \citep[e.g.][]{tempo2, wm1}. The TM accounts for all deterministic physical effects between pulse emission and reception. Therefore, timing residuals contain any unmodelled signals, including stochastic processes (for example noise sources) and potentially a GWB. Consequently, data analysis pipelines are designed to fit for both candidate GW signals and pulsar noise simultaneously \citep{nanograv_noise_budget, wm2}.
This simultaneous fit is particularly challenging, because pulsar observations are affected by numerous subtle stochastic noise sources, that are usually strongly correlated with the GW signal \citep[e.g.][]{2016MNRAS.455.4339T, 2018CQGra..35m3001V}. The three dominant sources of noise are: (i) white noise (WN), (ii) red noise (RN), which arises from random variations in pulsar spin rate  \citep[e.g.][]{2016MNRAS.458.2161L}, and (iii) dispersion measure (DM) noise, which is caused by fluctuations in the electron density of the interstellar medium along the line of sight \citep[e.g.][]{2004hpa..book.....L}. 
RN, DM and GWB are all time-correlated noise processes with red spectra, which can be distinguished through their distinctive signatures: the GWB is common to all pulsars, achromatic (radio frequency-independent) and exhibits a quadrupolar angular correlation \citep{hd1983}; RN varies from pulsar to pulsar, lacks spatial correlation and is achromatic; DM noise is unique to each pulsar and chromatic, scaling in amplitude as $\nu^{-2}$ with observing frequency \citep[e.g.][]{2004hpa..book.....L}. In principle, a PTA with many well-timed pulsars spread across the sky and observed over a wide frequency range can disentangle these contributions. In practice, however, current PTAs fall short of these ideal conditions. For example, the current evidence for GWB is dominated by only a few pulsars \citep[see][]{2023MNRAS.518.1802S}, and the frequency coverage, particularly in older legacy datasets, remains limited. Indeed, as demonstrated in \cite{2025A&A...694A..38F}, the scarce frequency coverage in the second data release of the EPTA strongly limits the evidence for a GWB signal. Quite surprisingly, in the real data this limitation appears to be overcome once the oldest data are removed. Indeed, the latest data release presented the results for two different versions of the EPTA \texttt{DR2}: \texttt{DR2full}, which contains all the backends and has a timespan of $\sim$25 years, and \texttt{DR2new}, which is limited to the last 10.3 years of data for all the pulsars. The result is that the GWB significance obtained with \texttt{DR2full} is $\lesssim 2\sigma$, while with \texttt{DR2new} it is $\gtrsim3.5\sigma$. This cut was made to exclude the mono-frequency measurements present in the first half of the data (the frequency band distribution of \texttt{DR2full} is displayed in the middle panel of \autoref{fig:SNRTobs}).\\

The main objective of this paper is to determine whether the counterintuitive increase in GWB significance obtained by cutting the data is an anomaly caused by the data itself, or whether it is indeed a possible outcome of the analysis pipeline.
We tackle this main question by simulating and analysing synthetic replicas of EPTA \texttt{DR2full} and \texttt{DR2new}. The paper is organized as follows. The simulation details and analysis techniques adopted are presented in Sec.~\ref{sec:sims}. The main results of our investigation are presented in Sec.~\ref{sec:res}. In particular in Sec.~\ref{sec:FC} we present detail results on the significance of GWB recovery in 100 mock realizations of \texttt{DR2full} and \texttt{DR2new}. In \ref{sec:PE}, we also evaluate the performance of  GWB parameter estimation in the two datasets, examining the effect of the spectral leakage resulting from the limited timespan of \texttt{DR2new}. This evaluation is conducted using the technique developed in \cite{2025arXiv250613866C} (Sec.~\ref{sec:FL}). In Sec.~\ref{sec:ipta}, we investigate how much the evaluation of the significance and the parameter estimation is improved by adding to the 25 pulsar of EPTA \texttt{DR2full} an IPTA-like frequency coverage. Finally, in section \ref{sec:5yr}, we study the behaviour of very short observation datasets (5yr of observation time), comparable to those recently presented by CPTA and MeerKAT, in the GWB parameter estimation. We summarize our conclusions in Sec.~\ref{sec:con}.

\section{Simulations and analysis tools}
\label{sec:sims}

\begin{table}[h]
\centering
\scriptsize 
\renewcommand{\arraystretch}{1.3}
\begin{tabularx}{\columnwidth}{|c|c|X|}
\hline
\textbf{Name} & \textbf{$T_{\rm obs}$} & \textbf{Description} \\
\hline
\texttt{DR2full} & 25yr & Same properties as the real EPTA \texttt{DR2full} \\
\hline
\texttt{DR2new} & 10yr & Last 10yr of \texttt{DR2full} simulations \\
\hline
\texttt{DR2 5yr} & 5yr & Last 5yr of \texttt{DR2full} simulations \\
\hline
\texttt{DR2* 5yr} & 5yr & Same properties as \texttt{DR2 5yr} with the amplitude of red noise and DM five times lower \\
\hline
\texttt{DR2FC 25yr} & 25yr & Same properties of \texttt{DR2full}, but with the frequency coverage homogeneous over $T_{\rm obs}$ \\
\hline
\texttt{DR2FC 10yr} & 10yr & Last 10yr of \texttt{DR2FC 25yr} \\
\hline
\texttt{DR2full+IPTA} & 25yr & \texttt{DR2full} + longest baseline backends from NANOGrav and PPTA + LOFAR + NenuFAR \\
\hline
\end{tabularx}
\caption{Simulated datasets used in this work.}
\label{tab:datasets}
\end{table}

\subsection{Reference datasets}

The reference datasets are the EPTA \texttt{\texttt{DR2full}} and EPTA \texttt{DR2new}, which are the 24.8yr and 10.3yr version of the EPTA \texttt{DR2}. \texttt{DR2new} is obtained from \texttt{DR2full} by removing the so-called \textit{legacy data}, recorded by the oldest backends. EPTA data are taken from the major European radio telescopes:  Effelsberg 100-m radio telescope (EFF) in Germany,  76-m Lovell Telescope, and  Mark II Telescope at Jodrell Bank Observatory (JBO) in the United Kingdom, along with the large radio telescope operated by the Nançay Radio Observatory (NRT) in France,  64-m Sardinia Radio Telescope (SRT) operated by the Italian National Institute for Astrophysics (INAF) through the Astronomical Observatory of Cagliari (OAC), and  Westerbork Synthesis Radio Telescope (WSRT) operated by ASTRON, the Netherlands Institute for Radio Astronomy. These telescopes also operate together as the Large European Array for Pulsars (LEAP), which gives an equivalent diameter of up to 194m \citep[]{2016MNRAS.460.2207B}.

\subsection{Simulation of a realistic PTA dataset}\label{sec:dr2_sim}

This section briefly describes the main steps involved in creating simulations which are as realistic as possible. For a more detailed explanation, see \cite{2025A&A...694A..38F}.

\subsubsection{Simulating realistic Times of Arrival}
\label{subsec:sim_toas}

The simulation pipeline begins by replicating the key characteristics of each pulsar’s observations, including observation times, number of observations, and cadence. The temporal variations in cadence and the large gaps present in some pulsars’ TOAs are also reproduced. Each simulated epoch is assigned the same radio frequency as in the corresponding real observation, along with the associated observatory, backend, and timing error. Data are then generated using the \texttt{libstempo} package \citep{2020ascl.soft02017V}, adopting the timing model (TM) from \cite{wm1}. The same package is then used to inject the noise budget, as described below.

\subsubsection{Simulating realistic noise budget}
\label{subsec:stat_gp_noise}

We model three noise components-- white noise, red noise, and dispersion measure\footnote{For a more in depth study, one should include also the solar wind noise components, which has been proven to affect the GWB search \citep{2016MNRAS.455.4339T, 2024A&A...692A..18S}} -- all treated as stationary Gaussian processes.
As such, these components are fully described by their covariance matrices. WN is uncorrelated in time, resulting in a diagonal covariance matrix with elements given by
\begin{equation}
C_{ii}^{\rm WN}=\sigma_i^2 = \mathrm{EFAC}_i^2 \, \sigma_{\mathrm{ToA}}^2 + \mathrm{EQUAD}_i^2,
\end{equation}
where $\sigma_{\mathrm{ToA}}$ is the template-fitting timing uncertainty, EFAC is a multiplicative factor applied to the ToA errors, and EQUAD accounts for additional backend-dependent white noise.

RN and DM variations, in contrast, are time-correlated and modeled in Fourier space by means of sine and cosine basis functions with normally distributed coefficients determined by their power spectral density $S(f)$ 
\begin{equation}
    S(f; A, \gamma) = \frac{A^2}{12\pi^2}\biggl(\frac{f}{{\rm yr}^{-1}}\biggr)^{-\gamma}{\rm yr}^3.
\end{equation}
Noise realizations are produced by randomly drawing Fourier coefficients around this power law $S(f)$, generating a covariance matrix that sums contributions from different frequencies and incorporates a chromatic index $i_d$, with $i_d = 0$ for achromatic RN and $i_d = 2$ for DM noise
\begin{equation}
\begin{aligned}
    C_{ij}^{\rm RN/DM} = \sum_k & S(f_k) \Delta f \bigg (\frac{\nu_{\rm ref}}{\nu_i} \bigg )^{i_{d}} \bigg(\frac{\nu_{\rm ref}}{\nu_j}  \bigg )^{i_{d}} \cdot \cos(2\pi f_k (t_i - t_j))
\end{aligned}
\label{eq:cov}
\end{equation}
where $\nu_i$ is the observed radio frequency at TOA $t_i$. The amplitude and slope used in the $S(f)$ for the injection are the maximum likelihood values obtained from the single pulsar noise analysis (SPNA) on EPTA \texttt{DR2} real data \citep{wm2}. The number of Fourier components $f_k = k / T_{\rm obs}$ used for the injection is also taken from the SPNA. This approach ensures that the simulated timing data reproduce the statistical characteristics of real pulsar observations.

\subsubsection{Simulating a realistic GWB}
\begin{figure}
   \centering
   \includegraphics[width=0.5\textwidth]{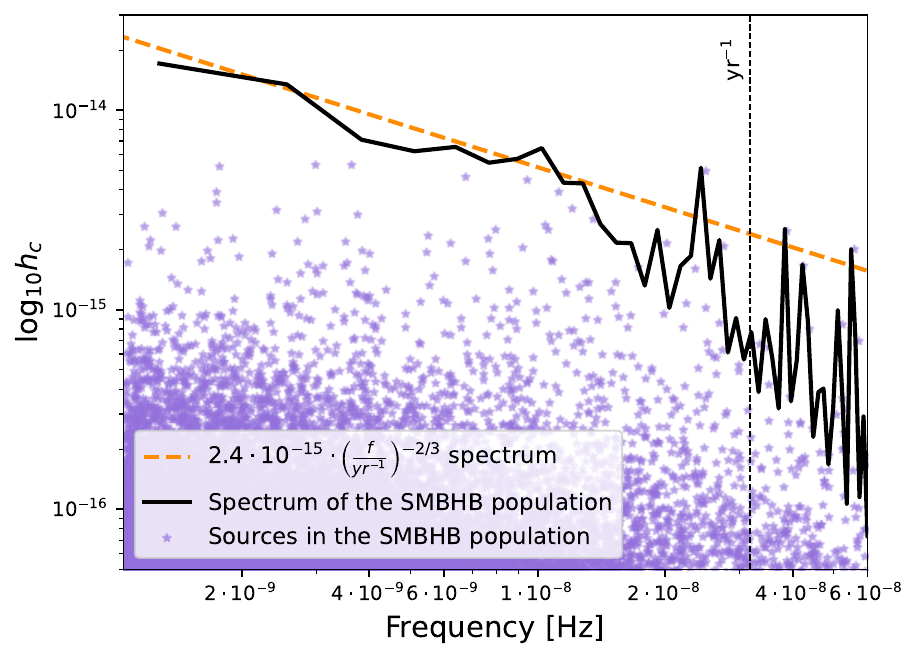}
   \caption{ Characteristic strain spectrum of the injected GWB. Purple stars are the contributions to the spectrum of all the SMBHBs in the populations. The black solid line is the total power spectrum in each frequency bin, where the frequency bins start at $f = 1/T_{\rm obs}$ and have width $\Delta f = 1/T_{\rm obs}$ with $T_{\rm obs}$ = 24.8yr. The power law approximation of the spectrum is also shown.}
   \label{fig:spectrum}%
\end{figure}

To simulate a GWB from SMBHBs, the injection pipeline generates a population of circular binaries, calculates the timing variations each induces on pulsar TOAs, and sums these delays. The population is sampled from the distribution $\mathrm{d}^3N/(\mathrm{d}z\,\mathrm{d}M\,\mathrm{d}f_r)$, which is the number of emitting sources per unit redshift, mass and frequency and is based on observation-driven models \citep{sesana_2013b, Rosado_2015}. 
The power spectrum of such a population is 
\begin{equation}
h_c^2(f_i) = \sum_{j \in \Delta f_i} h_j^2(f_r)\frac{f_r}{\Delta f_i},
\end{equation}
where $f_i$ is the bin’s central frequency and $h_j(f_r)$ is the strain of each source, given by 
\begin{equation}
h(f_r) = 2 \sqrt{\frac{a(i)^2 + b(i)^2}{2}} \frac{(G\mathcal{M})^{5/3}(\pi f_r)^{2/3}}{c^4 d(z)},
\end{equation}
with $a(i)=1+\cos^2i$, $b(i)=-2\cos i$, and $d(z)$ the comoving distance:
\begin{equation}
d(z) = \frac{c}{H_0}\int_0^z \frac{1}{\sqrt{\Omega_m(1+z')^3 + \Omega_\Lambda}}\,dz'.
\end{equation}
In the time domain, the GW signal emitted by an inspiraling SMBHB is described by eight parameters: chirp mass $\mathcal{M}$, GW frequency $f_r$, redshift $z$, inclination $i$, polarization $\psi$, phase $\Phi_0$, and sky position $(\theta,\phi)$. $\mathcal{M}$, $f_r$, and $z$ are drawn from $\mathrm{d}^3N/(\mathrm{d}z\,\mathrm{d}\mathcal{M}\,\mathrm{d}f_r)$, while the remaining angular parameters are drawn such that the source sky location $(\theta,\phi)$ is random on the sphere, $i$ is randomly oriented on the sphere, $\psi$ and $\Phi_0$ are uniform in the ranges (0, $\pi$) and (0, 2$\pi$) respectively. Once the eight parameters of each binary in the SMBHB population are generated, the corresponding timing variations (accounting for Earth and pulsar terms) are computed following \citet{Ellis_2013} and summed in the time domain to yield the total GWB signal. \\
For this study, a single population of circular SMBHBs was used, which generates a power spectrum with $A_{\rm GWB} = 2.4\times10^{-15}$ when $\alpha_{\rm GWB}$ is fixed at -2/3. These values are consistent with the constraints from EPTA \texttt{DR2} \citep{wm3}.
The total spectrum together with the contribution of each binary in the population and the power law approximation is displayed in \autoref{fig:spectrum}. A schematic summary of all the simulated datasets used and their description can be found in \autoref{tab:datasets}.

\subsection{Data analysis}
We use standard PTA analysis techniques, based on the assumption of a Gaussian, stationary noise and adopt a Gaussian likelihood. Following \citet{vv2014}, first-order timing model (TM) parameter uncertainties are analytically marginalised, yielding the TM-marginalized likelihood:
\begin{equation}
\ln \mathcal{L}(\delta t|\theta) \propto -\frac{1}{2}\delta t^\top C^{-1} \delta t - \frac{1}{2} \ln \det C,
\end{equation}
where $\delta t = \sum_{a=1}^{N_\mathrm{PSR}} \delta t_a$ and the covariance matrix is $C = C_a \delta_{ab} + C_h \Gamma_{ab}$. Here, $C_h$ describes an achromatic common red noise (RN) correlated between pulsars via the Hellings–Downs curve \citep{hd1983}:
\begin{equation}
\Gamma_{ab} = \frac{1}{3} - \frac{1}{6} \frac{1-\cos \gamma_{ab}}{2} + \frac{1-\cos \gamma_{ab}}{2} \ln \left( \frac{1-\cos \gamma_{ab}}{2} \right),
\end{equation}
with $\gamma_{ab}$ the angular separation between the pair of pulsars $a$ and $b$. The power spectral density of this common red noise process is modelled as a power law, which is directly related to the GWB strain via
\begin{equation}
    S(f) = \frac{h_c^2(f)}{12\pi^2f^3} = \frac{A_{yr}^2}{12\pi^2}\left( \frac{f}{1yr^{-1}} \right)^{-\gamma}\cdot{\rm yr}^{3} 
\end{equation}
where $\gamma = 3 - 2\alpha$.
The number of components used in the common red noise model is set such that the maximum frequency is the highest frequency bin below $f=1$yr$^{-1}$ (this means that for the 25yr datasets the number of components used is 24, for the 10yr is 9 and for the 5yr is 4). The term $C_a$ represents the noise in pulsar $a$:
\begin{equation}
C_a(t_i, t_j) = \sigma_i^2 \delta_{ij} + C^\mathrm{RN}(t_i-t_j) + C^\mathrm{DM}(t_i-t_j),
\end{equation}
where $\sigma_i$ is the white noise level at $t_i$, and $C^\mathrm{RN}$ and $C^\mathrm{DM}$ are red noise and dispersion measure covariance matrices given by \autoref{eq:cov}. Rather than assuming a power-law spectrum, the PSD in each frequency bin $k/T_{\rm obs}$ of width $\Delta f=1/T_{\rm obs}$ can be treated as a free parameter. This analysis is commonly referred to as free spectrum analysis.

Given the likelihood, the analysis is run in a Bayesian framework, where the posterior probability distribution of the model parameters $\vec{\theta}$ is obtained by updating our prior knowledge $\pi(\vec{\theta})$ with the information provided by the data. This is done by multiplying the prior with the likelihood function $\mathcal{L}(\delta t|\vec{\theta})$, which quantifies the probability of the observed timing data $\delta t$ given $\vec{\theta}$. The resulting posterior probability is  
\begin{equation}
    p(\vec{\theta}|\delta t) = 
    \frac{\mathcal{L}(\delta t|\vec{\theta})\,\pi(\vec{\theta})}
    {\int \mathcal{L}(\delta t|\vec{\theta})\,\pi(\vec{\theta})\,d\theta},
\end{equation}
which expresses the probability of the parameters $\vec{\theta}$ conditioned on the timing data $\delta t$. The denominator is the normalisation factor known as the \emph{Bayesian evidence}. To estimate the posterior distribution from the simulated data, we use the likelihood model implemented in \texttt{ENTERPRISE\_EXTENSIONS}, sampled with a Markov Chain Monte Carlo (MCMC) algorithm.  To evaluate the significance of the GWB signal, we use the noise-marginalised optimal statistic described in \cite{2018PhRvD..98d4003V}, which is a commonly employed hybrid of Bayesian and frequentist techniques. 

\section{Results}
\label{sec:res}

\subsection{Understanding \texttt{DR2full} and \texttt{DR2new} results}

In this section, we study the behaviour of the \texttt{DR2full} and \texttt{DR2new} simulations and compare them to the real \texttt{DR2full} and \texttt{DR2new} datasets. To this end, we simulate 100 realisations of a 24.8-year dataset that closely resembles the real \texttt{DR2full} one, as described in \autoref{sec:dr2_sim}. These 100 realisations are obtained by injecting the same GWB signal (whose spectrum is shown in \autoref{fig:spectrum}) each time with a different statistical realisation of the noise. Shorter versions of \texttt{DR2} are then obtained by selecting the TOAs from a chosen epoch to the most recent TOA, such that the cut at 10yr corresponds to \texttt{DR2new}. Since the impact of the frequency coverage appears crucial for understanding the behaviour of \texttt{DR2full} and \texttt{DR2new}, we have also created a version of \texttt{DR2} with the same observation epochs but homogeneous coverage of observation frequencies throughout the entire observation time $T_{\rm obs}$. This second version of the dataset is referred to as \texttt{DR2FC}, where FC stands for frequency coverage. The difference in frequency coverage between the two datasets is displayed in \autoref{fig:SNRTobs}. \texttt{DR2FC} contains the same GWB as well as the same 100 noise realisations that were injected into \texttt{DR2}.

\subsubsection{GWB significance and the impact of the frequency coverage}
\label{sec:FC}

\begin{figure}[h]
   \centering
   \includegraphics[width=0.5\textwidth]{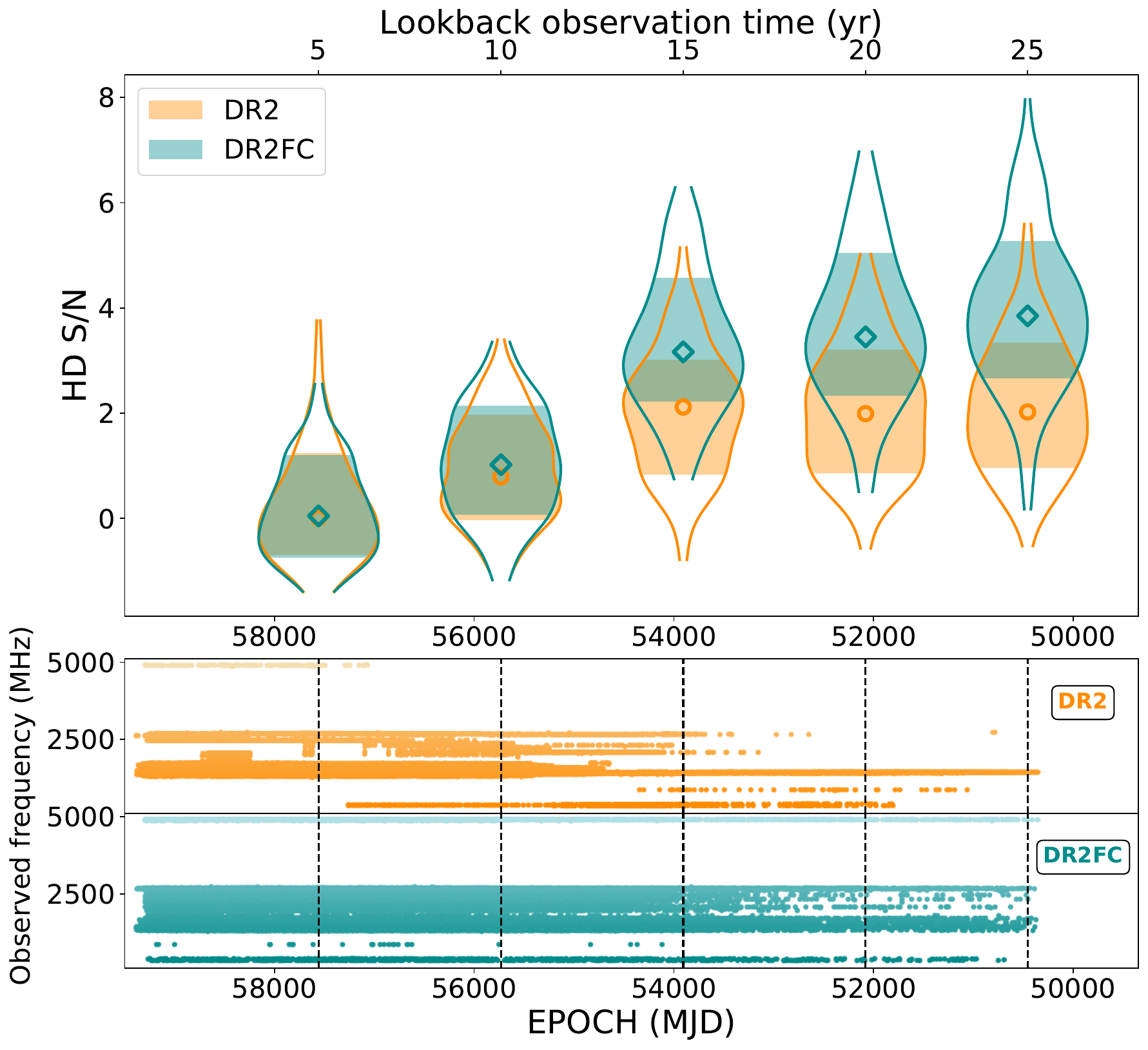}
   \caption{Top: Distribution of the SNR computed from the 100 realizations of \texttt{DR2} (orange) and \texttt{DR2FC} (green). Bottom: Difference in frequency coverage between the two kinds of datasets. The TOAs from all the pulsars are plotted together to highlight the drop in the frequency coverage occurring in \texttt{DR2} at 15yr from now, which corresponds to the plateau of its SNR.}
   \label{fig:SNRTobs}%
\end{figure}
\begin{figure*}
    \centering
    \begin{subfigure}{0.48\textwidth}
        \centering
        \includegraphics[width=\linewidth]{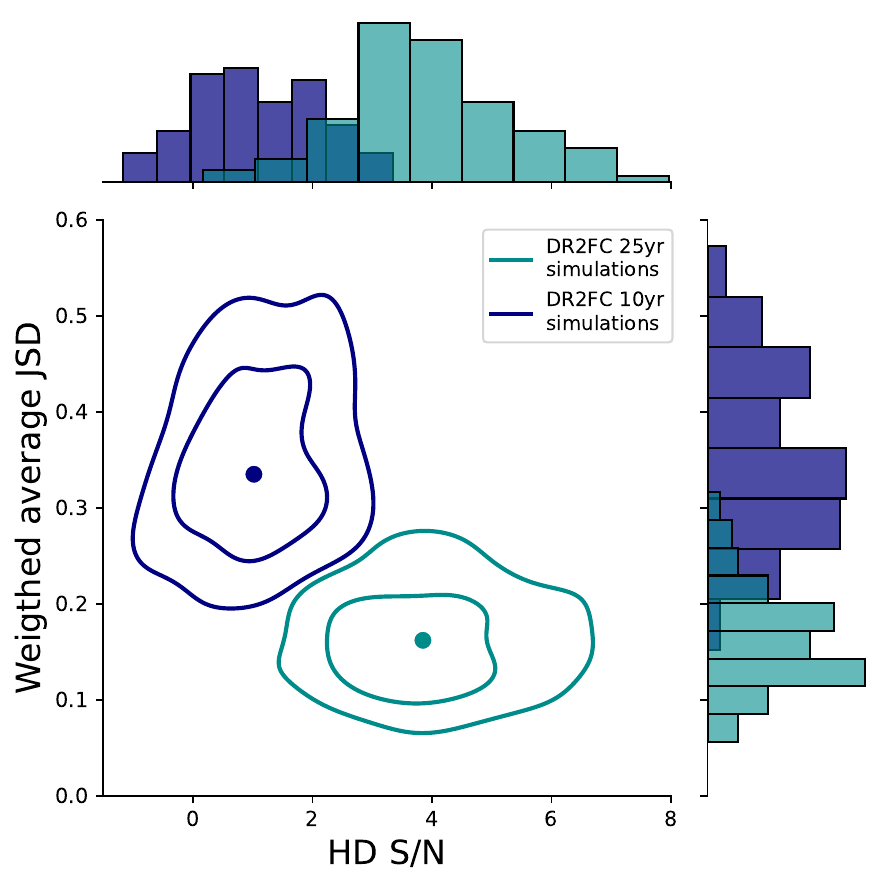}
        \caption{Distribution for the 25yr and the 10yr version of \texttt{DR2FC}, which is EPTA \texttt{DR2} with a homogeneous frequency coverage over the entire observation time.}
        \label{fig:SNR_JSD_DR2fc}
    \end{subfigure}
    \hfill
    \begin{subfigure}{0.48\textwidth}
        \centering
        \includegraphics[width=\linewidth]{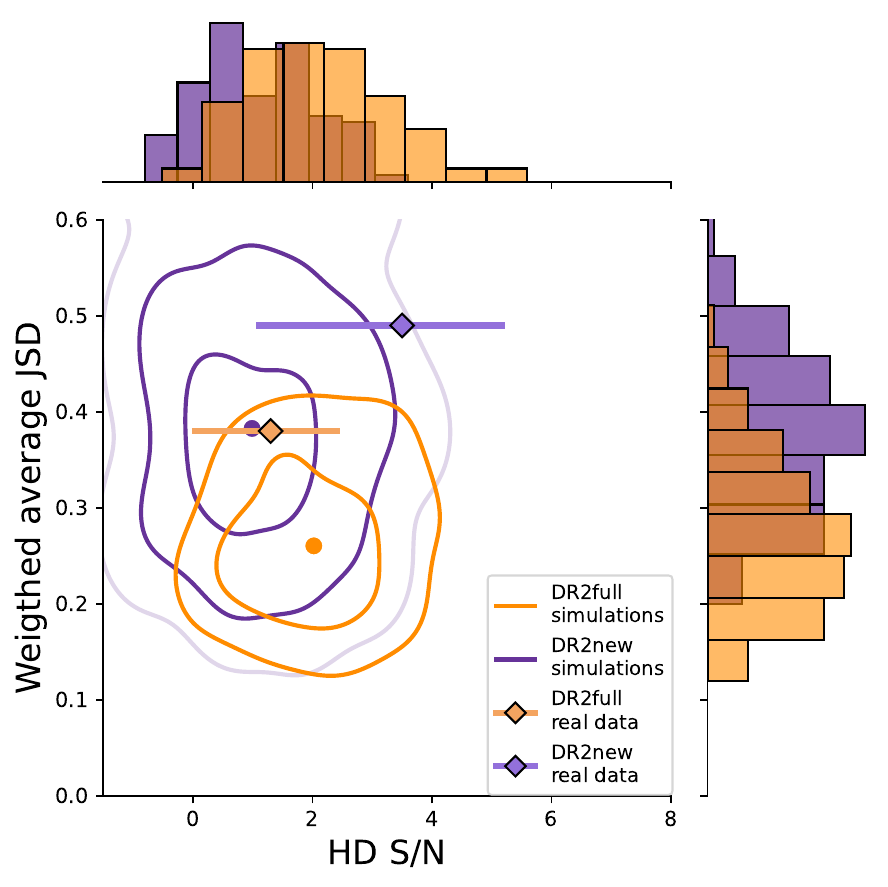}
        \caption{Distribution for the 25yr (\texttt{DR2full}) and the 10yr (DR2new) version of \texttt{DR2}. As a comparison, the values of JSD and HD S/N with 90$\%$C.I. computed from the real data.}
        \label{fig:SNR_JSD_DR2}
    \end{subfigure}
    \caption{2-d distributions of the noise marginalised HD S/N and the degeneracy coefficient (weighted average JSD)}
    \label{fig:twosubfigs}
\end{figure*}

\autoref{fig:SNRTobs} shows the distribution of HD S/N for each of the 100 realisations of the \texttt{DR2} and \texttt{DR2FC} datasets as a function of observation time. To create this plot, each \texttt{DR2} and \texttt{DR2FC} realisation was cut at 5, 10, 15 and 20 years from the last TOA. A Bayesian search for a common uncorrelated red noise signal (CURN) was then performed on each of these realisations, with all noise components left as free parameters. The value of the HD S/N was finally obtained by evaluating the median of the HD S/N distribution obtained using the noise marginalised optimal statistics method. As can be seen, the average behaviour of the two datasets is extremely similar up to an observation time of 10 years (\texttt{DR2new}), after which they diverge. For the \texttt{DR2FC} dataset, the S/N continues to grow, although a slowdown in growth is evident after 15 years, which is related to the decrease in the number of data points and the increase of the TOA errors in the early 5-year data segments. The S/N growth in \texttt{DR2} slows down significantly after 10 years, compared to \texttt{DR2FC}, and stops for observation times greater than 15 years: the first ten years of data, on average, do not contribute to the HD S/N. This is due to the lack of frequency coverage in the first ten years of \texttt{DR2}: while the number of frequency channels remains significant up to 15 years, beyond this point, individual pulsars data feature a single frequency channel.  Then the legacy data are unable to resolve the degeneracy between the frequency-dependent DM noise model and the achromatic GWB model. Indeed, if only one frequency channel is available, the DM noise model is indistinguishable from an achromatic noise model (see \autoref{eq:cov}). The average behaviour shows that the first 10 years of \texttt{DR2} are not informative enough to contribute to the GWB significance compared to the last 15 years.\\
The reason why the scarce frequency coverage implies a low GWB significance was investigated in \cite{2025A&A...694A..38F} and is related to the degeneracy between the GWB parameters and the individual pulsar noise parameters. This degeneracy is quantified by a degeneracy coefficient, which assigns a significance to the tails of the posterior distributions using the Jensen–Shannon divergence (JSD) with respect to a Gaussian distribution. The JSD is evaluated for all 31 noise processes and then averaged to obtain a single coefficient that quantifies the level of degeneracy for each realisation \citep[the {\it weighted average JSD}, see][for a detailed description of this stastistic]{2025A&A...694A..38F}.
The relation between the GWB significance and the degeneracy coefficient is illustrated in \autoref{fig:SNR_JSD_DR2fc} and \autoref{fig:SNR_JSD_DR2}. Here, we selected the HD S/N and weighted average JSD values for the 25-year (full) and 10-year (new) versions of the datasets. \autoref{fig:SNR_JSD_DR2fc} shows that moving from the 10-year to the 25-year \texttt{DR2FC} datasets significantly improves the ability to distinguish between individual noise and GWB components, decreasing the mean JSD. Consequently, the HD S/N increases significantly, rising from a median value of $\sim$ 1.2 for the 10-year \texttt{DR2FC} dataset to $\sim$ 3.8 for the 25-year \texttt{DR2FC} dataset. In the \texttt{DR2} dataset, however, the decrease in the JSD is much smaller, as is the increase in the HD S/N, which goes from $\sim$ 1.0 (\texttt{DR2new}) to $\sim$ 2.0 (\texttt{DR2full}). This causes the HD S/N distributions for \texttt{DR2new} and \texttt{DR2full} to largely overlap. The consequence of this overlap is that, because of random fluctuations in the value of the HD S/N, \texttt{DR2new} can give a higher GWB significance than \texttt{DR2full} with the same noise realisation: this happens in the 15$\%$ of the realisations. In this 15$\%$, the average increase in the HD S/N from \texttt{DR2full} to \texttt{DR2new} is of order 50$\%$. On the other side, there is only a 2$\%$ probability of having a higher significance from \texttt{DR2FC} 10yr than from \texttt{DR2FC} 25yr, because of the much smaller overlap between the two distributions.\\
To see whether EPTA observations are compatible with statistical fluctuations found in our simulations, we selected the realisations in which \texttt{DR2new} gives a higher S/N than \texttt{DR2full}, and the medians of the S/N from the simulations are compatible within 90$\%$C.I. with the HD S/N values obtained from the data. Five realisations were selected using this criterion: the probability of a scenario similar to that observed in EPTA \texttt{DR2} occurring in our controlled set of simulations is 5$\%$, which is consistent with it being a $\approx2\sigma$ random fluctuation due to the noise realisation. However, if we repeat the same exercise with the \texttt{DR2FC} dataset, none of the realisations behave similarly to the real data. This confirms that the sparse frequency coverage in the first half of the data is primarily responsible for this behaviour. In the five selected cases, adding the first 15 years of mainly mono-frequency data to \texttt{DR2new} degrades the GWB significance instead of improving it. The values of the HD S/N from these cases are displayed in \autoref{tab:specials}, and an example of the HD S/N evolution with $T_{\rm obs}$ is shown in \autoref{fig:94_snr}. We will explore them more closely in the next section\\ 
\begin{figure*}[h]
    \centering
    \begin{subfigure}[t]{0.32\textwidth} 
        \centering
        \includegraphics[width=\linewidth]{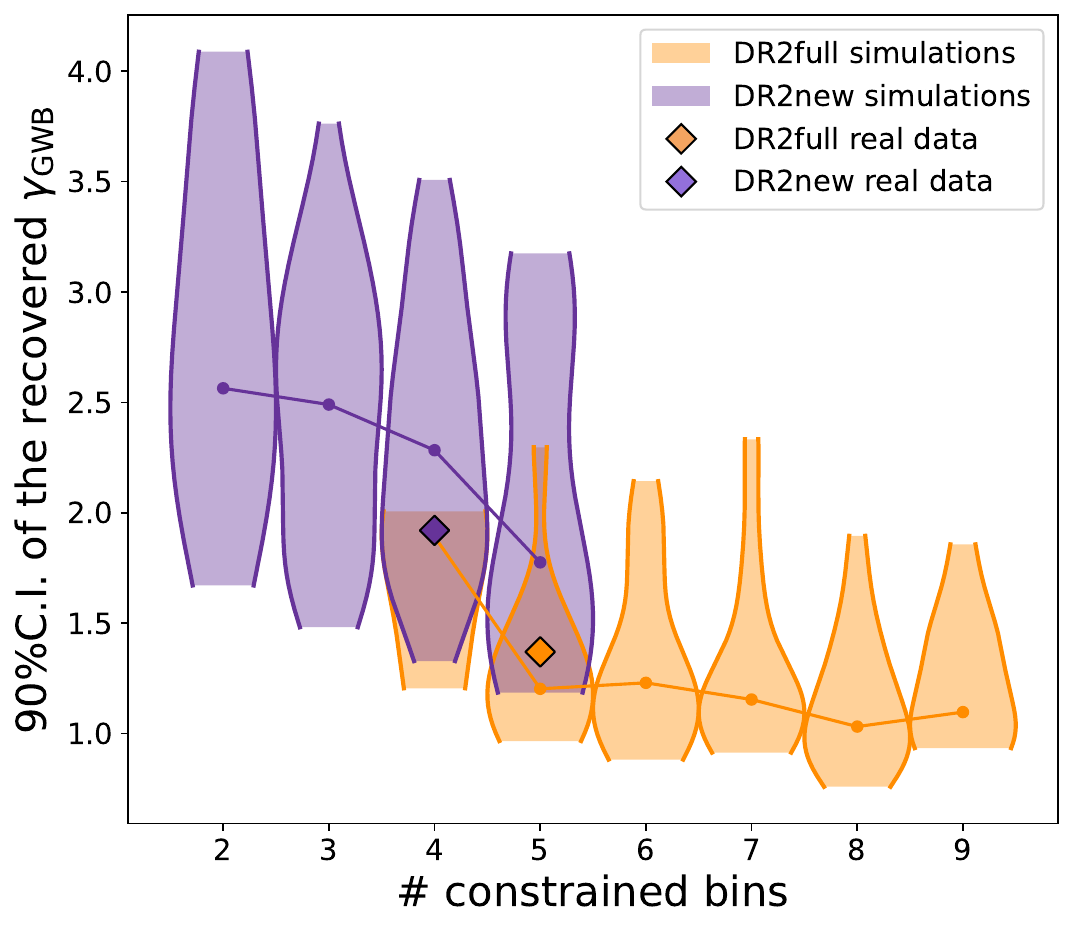}
        \caption{Distribution of the 90$\%$C.I. width of the $\gamma_{\rm GWB}$ posterior distribution as a function of the number of frequency bins that are constrained in the freespectrum search.}
        \label{fig:dG}
    \end{subfigure}
    \hspace{0.01\textwidth} 
    \begin{subfigure}[t]{0.32\textwidth}
        \centering
        \includegraphics[width=\linewidth]{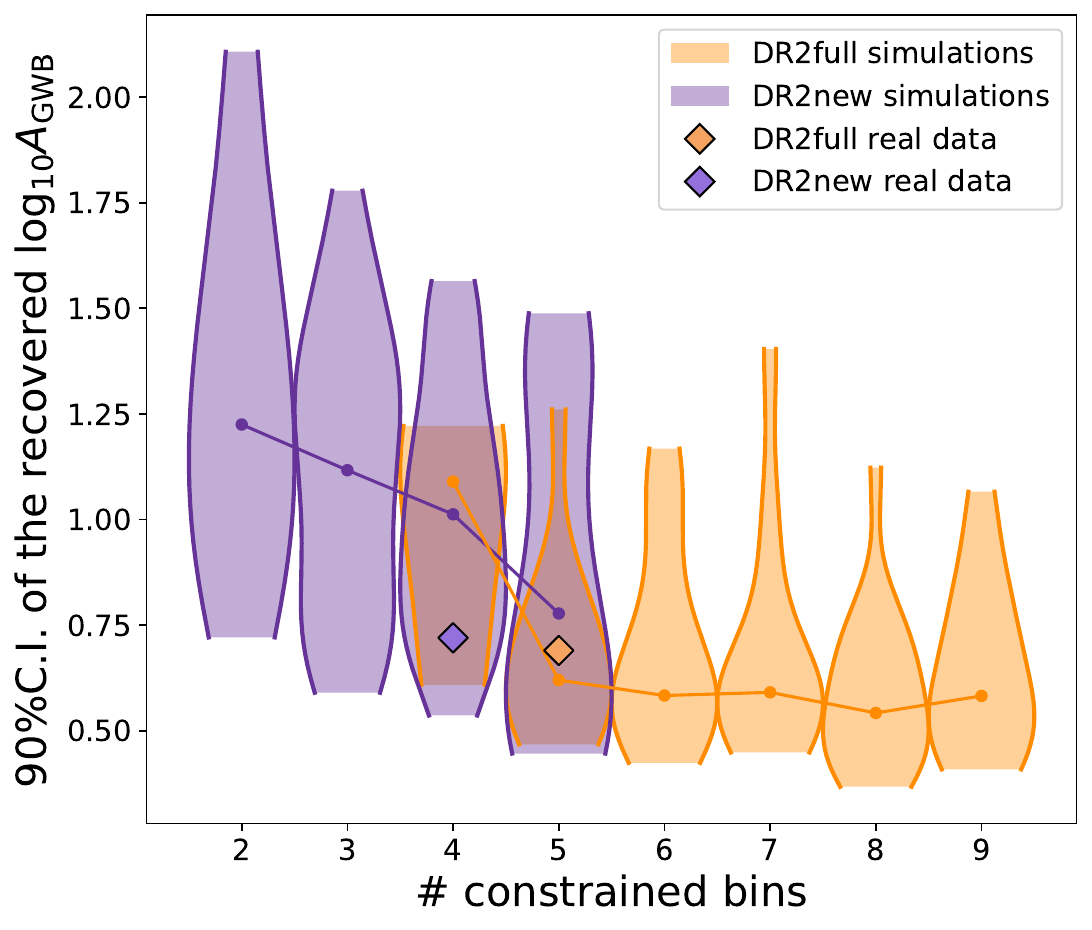}
        \caption{Distribution of the 90$\%$C.I. width of the log$_{10}A_{\rm GWB}$ posterior distribution as a function of the number of frequency bins that are constrained in the freespectrum search.}
        \label{fig:dA}
    \end{subfigure}
    \hspace{0.01\textwidth} 
    \begin{subfigure}[t]{0.32\textwidth}
        \centering
        \includegraphics[width=\linewidth]{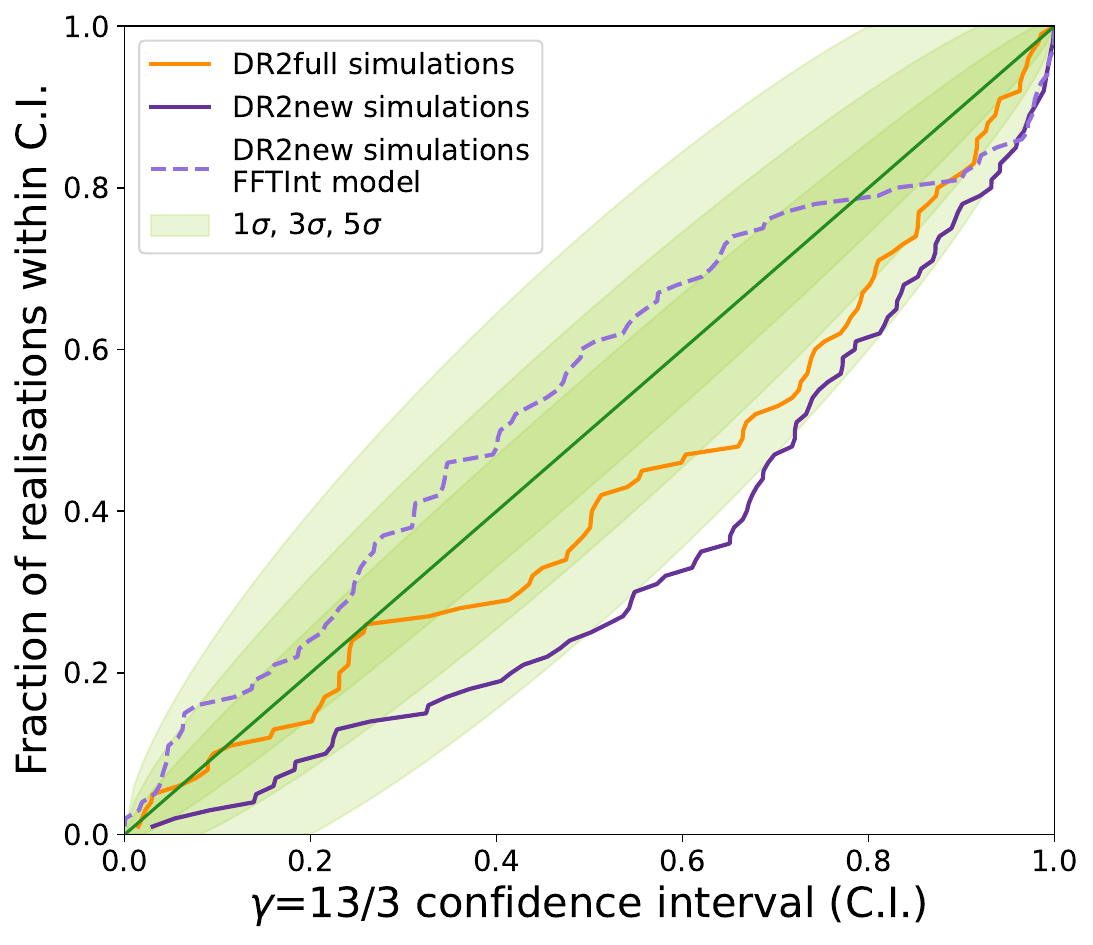}
        \caption{P-P plot computed with the marginalised posterior distribution of $\gamma_{\rm GWB}$, taking as the true value the standard $13/3$. The green diagonal represents the true shape of the P-P plot for a perfectly unbiased recovery, the shaded areas are the 1, 3 and 5$\sigma$ C.I. given the number of data, which is 100.}
        \label{fig:PP_plot}
    \end{subfigure}
    \caption{Comparison of the precision and accuracy of the GWB parameter estimation performed with \texttt{DR2full} (orange) and \texttt{DR2new} (purple).}
    \label{fig:twosubfigs}
\end{figure*}

Finally, it should be noted that the HD S/N values estimated from the simulations are consistent with those obtained from real data. The EPTA \texttt{DR2} values are shown in \autoref{fig:SNR_JSD_DR2},  together with the corresponding weighted average JSD values. The \texttt{DR2full} simulations agree with the median of the HD S/N computed from the real data within the 68$\%$ C.I.: this agreement was also observed with another set of 100 noise realisations in \cite{2025A&A...694A..38F}. \texttt{DR2new} shows worse agreement, but the overlap between the HD S/N distribution obtained from the simulations and the 90$\%$C.I. estimated from the real data  \citep[S/N$_{\rm HD} = 3.5^{+2.4}_{-1.7}$,][] {wm3} is significant.
This shows that the behaviour observed in the real data is unlikely, but cannot be ruled out by simulations.

\subsubsection{GWB parameter estimation with \texttt{DR2full} and \texttt{DR2new}}
\label{sec:PE}

\begin{figure}
   \centering
   \includegraphics[width=0.5\textwidth]{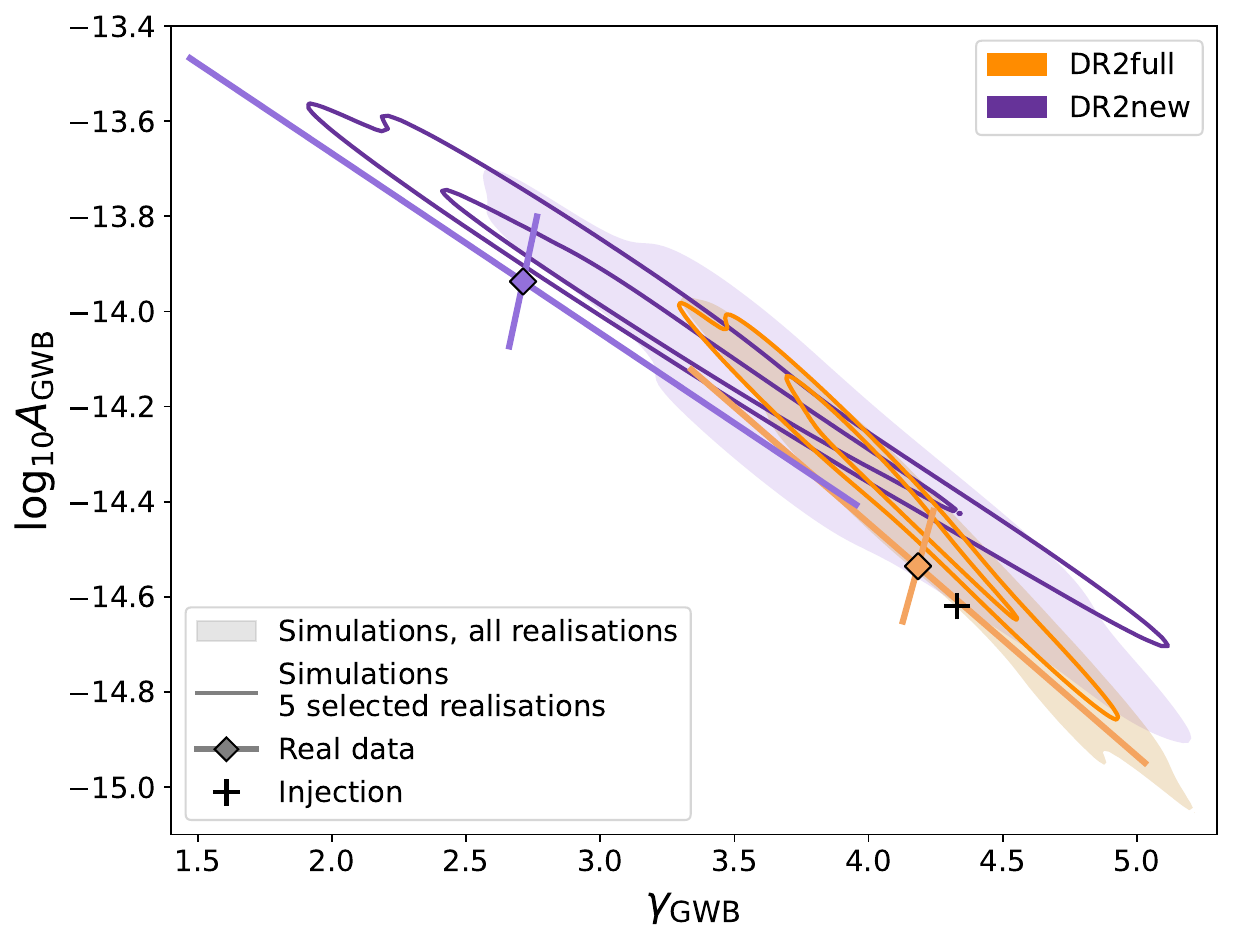}
   \caption{Median of the marginalised posterior distributions of $\gamma_{\rm GWB}$ and $\log_{10}A_{\rm GWB}$ of the 100 realisations of \texttt{DR2full} and \texttt{DR2new} (shaded regions, 95$\%$C.I.) and of the 5 realisations selected on the HD S/N (solid lines, 68 and 95$\%$C.I.). As a reference, the median values and 2$\sigma$ covariance principal axis obtained from the real data are displayed.}
   \label{fig:PE_selected}%
\end{figure}
\begin{table*}[h]
\centering
\begin{tabular}{c|ccc|cc|cc|ccc}
\toprule
\multicolumn{1}{c|}{\textbf{Realisation}} & \multicolumn{3}{c|}{\textbf{HD S/N}} & \multicolumn{2}{c|}{\textbf{(log$_{10}A_{\rm GWB}$, $\gamma_{\rm GWB}$)}} & \multicolumn{2}{c|}{\textbf{13/3 C.I.}} & \multicolumn{3}{c}{\textbf{JSD}}\\
\cmidrule(lr){1-1}\cmidrule(lr){2-4} \cmidrule(lr){5-6} \cmidrule(lr){7-8} \cmidrule(lr){9-11}
 & \texttt{DR2full} & \texttt{DR2new} & \texttt{IPTA} & \texttt{DR2full} & \texttt{DR2new} & \texttt{DR2full} & \texttt{DR2new} & \texttt{DR2full} & \texttt{DR2new} & \texttt{IPTA} \\
\midrule
17 & 2.1 & 2.4 & 4.6 & (-14.2, 3.7) & (-14.2, 3.5) & 97$\%$ & 81$\%$ & 0.30 & 0.46 & 0.15 \\
61 & 2.2 & 2.5 & 4.4 & (-14.3, 4.0) & (-14.4, 4.4) & 86$\%$ & 45$\%$ & 0.27 & 0.35 & 0.39 \\
\rowcolor{orange!20}
69 & 2.1 & 2.9 & 3.1 & (-14.2, 3.9) & (-14.0, 3.1) & 72$\%$ & 99$\%$ & 0.16 & 0.25 & 0.15 \\
\rowcolor{orange!20}
70 & 1.4 & 2.3 & 4.1 & (-14.7, 4.6) & (-13.8, 2.7) & 75$\%$ & 99$\%$ & 0.18 & 0.40 & 0.17 \\
\rowcolor{orange!50}
94 & 1.9 & 2.3 & 3.2 & (-14.4, 4.1) & (-13.9, 2.9) & 70$\%$ & 98$\%$ & 0.36 & 0.60 & 0.26 \\
\bottomrule
Real data & 1.3 & 3.5 & & (-14.5, 4.2) & (-13.9, 2.7) & 64$\%$ & 97$\%$ & 0.38 & 0.49\\
\end{tabular}
\caption{Comparison of the main outputs of the GWB analysis obtained with 5 simulated dataset (selected according to the agreement with the HD S/N of the real EPTA \texttt{DR2}) and the real data. The main outputs are the median of the noise marginalised optimal statistics HD S/N, the median of the marginalised posterior distribution of the slope and the ampliutde of the GWB, the confidence interval corresponding to the nominal value $\gamma_{\rm GWB} = 13/3$ and the degeneracy coefficient (JSD). The HD S/N and the JSD values are shown also for the synthetic dataset \texttt{DR2full + IPTA} investigated in \autoref{sec:ipta}.}
\label{tab:specials}
\end{table*}

As with the GWB significance, the GWB parameter estimation is sensitive to observation time. In this section, we discuss why the parameter estimation obtained from \texttt{DR2new} is less precise and less accurate than that obtained from \texttt{DR2full}.
All results presented in this section are referred to the standard analysis performed in the EPTA \texttt{DR2} release papers \citep{wm3}, upon which our investigation is based. Further effects can contribute to parameter estimation bias, as pointed out in \cite{goncharov2025readingsignaturessupermassivebinary}. We will discuss those in more detail towards the end of this section.
\\

\texttt{DR2full} is expected to give a more precise estimate than \texttt{DR2new} because a longer baseline gives access to more frequency components and a lower minimum frequency. Indeed, the 90$\%$C.I. on $\gamma_{\rm GWB}$ and log$_{10}A_{\rm GWB}$ obtained from the simulations are ($1.2^{+0.4}_{-0.2}, 0.6^{+0.3}_{-0.1}$) for \texttt{DR2full} and ($2.3^{+0.7}_{-0.7}, 1.0^{+0.4}_{-0.4}$) for \texttt{DR2new}, so the improvement is approximately 50$\%$ for the slope and 40$\%$ for the amplitude. This is in good agreement with the 90\%C.I. we have in the data: (1.37, 0.69) for \texttt{DR2full} and (1.92, 0.72) for \texttt{DR2new}. In both the data and the simulations, the improvement in precision is larger for $\gamma_{\rm GWB}$ than for log$_{10}A_{\rm GWB}$. This is expected, because the error on the slope depends more on the observation time than the error on the amplitude, as discussed in \autoref{sec:5yr} and demonstrated in \autoref{appendix:B}.\\

Furthermore, we can notice from \autoref{fig:dG} and \autoref{fig:dA} a dependence of the precision on the number of frequency bins that the data are able to constrain: the more frequency bins are constrained, the smaller is the error on the GWB parameters. Here, a free spectrum analysis is run and whether a frequency bin is constrained or not is decided with the Bayes Factor approximated by the Savage–Dickey ratio (i.e. the ratio of the prior and posterior distributions evaluated at zero or, in this case, at the lower edge of the prior distribution). If the Bayes factor is bigger than 1, the frequency bin is considered as constrained. In the real data, the number of constrained bins is five for \texttt{DR2full} and four for \texttt{DR2new}. The similar number of constrained bins can be the reason why the improvement in the precision is smaller in the data than it is, on average, in the simulations (respectively 28$\%$ and 5$\%$ against an average of 50$\%$ 40$\%$ for $\gamma_{\rm GWB}$ and log$_{10}A_{\rm GWB}$). Still, some of the constrained bins in the real data might come from unmodelled noise sources that are degenerate with the GWB. This should be determined by in-depth noise analysis campaigns.

Turning now to the accuracy of the GWB parameter estimation, we built a probability-probability (P-P) plot in order to analyse the performance of \texttt{DR2full} and \texttt{DR2new} in providing an unbiased estimate of the GWB parameters. Specifically, we are interested in whether they can recover a GWB spectrum with $\gamma=13/3$. The P-P plot shows on the $y$-axis the fraction of times in which the nominal injection value lies within the credible interval indicated on the $x$-axis. In the case of unbiased and confident inference, the data points follow the diagonal of the plot parameter space. 
As can be seen from the P-P plot in \autoref{fig:PP_plot}, the recovery obtained from \texttt{DR2new} is more biased than that obtained from \texttt{DR2full}, preferring lower values of $\gamma_{\rm GWB}$ (and consequently higher values of the amplitude, see \autoref{fig:PE_selected}). The difference is quite large: the significance of the bias is $\sim3\sigma$ for \texttt{DR2full} and $\geq5\sigma$ for \texttt{DR2new}. It must be noticed that a perfect agreement with $\gamma=13/3$ is not to be expected even for a flawless PTA, because of the mismatch between the recovery model, a perfectly Gaussian power-law shaped GWB, and the injected spectrum (see \autoref{fig:spectrum}), as stated in \cite{2024A&A...683A.201V}.
The bias in both datasets is always towards a flatter and higher GWB spectrum, as can be seen from \autoref{fig:PE_selected}. Indeed, the median and 90$\%$C.I. of the 100 realisations are respectively ($4.17^{+0.78}_{-0.66}$, $-14.43^{+0.24}_{-0.58}$) for \texttt{DR2full} and ($3.92^{+1.47}_{-0.99}$, $-14.31^{+0.43}_{-0.66}$) for \texttt{DR2new}. The same behaviour -- which is the recovered GWB spectrum being significantly flatter in \texttt{DR2new} than in \texttt{DR2full} -- is observed in the real data, where the measured GWB parameters from \texttt{DR2full} are $(4.19^{+0.73}_{-0.63}, -14.54^{+0.28}_{-0.41})$ and $(2.71^{+1.18}_{-0.71}, -13.94^{+0.23}_{-0.48})$ from \texttt{DR2new}. Therefore, on average, the behaviour of real data is qualitatively reproduced, even if the GWB estimates from the \texttt{DR2new} simulations remain systematically lower in amplitude and steeper in the shape than the measured spectrum. This tension is mitigated if we look at the subsample of simulations that were selected in the previous section as the most similar to the data: as can be seen from \autoref{fig:PE_selected}, in this case the distribution of recovered parameters stays approximately the same for \texttt{DR2full} ($\gamma_{\rm GWB}=3.98^{+0.50}_{-0.27}$, $\log_{10}A_{\rm GWB}=-14.30^{+0.11}_{-0.31}$), but it moves towards higher and flatter spectra for \texttt{DR2new} ($\gamma_{\rm GWB}=3.14^{+1.07}_{-0.43}$, $\log_{10}A_{\rm GWB}=-13.99^{+0.15}_{-0.39}$). Indeed, three of the five selected realisations behave very similarly to the real data in terms of parameter estimation (these are realisations 69, 70 and 94, see \autoref{tab:specials}): the recovered parameters from \texttt{DR2new} are $\gamma\lesssim3$ and $\log_{10}A\gtrsim-14$ and the agreement with the nominal $\gamma=13/3$ moves from within 1$\sigma$ C.I. for \texttt{DR2full} to between 2 and 3$\sigma$ for \texttt{DR2new}. The same occurs in the real data (see \autoref{tab:specials}). Obviously, we do not know the true parameters of the signal that is present in the data and $\gamma=13/3$ is only taken as a reference value. In the simulations, this increase in the bias from \texttt{DR2full} to \texttt{DR2new} is related to an increase in the degeneracy between individual noise and the GWB parameters (see again \autoref{tab:specials}). The same increase is observed in the posterior distributions of the real data. This degeneracy is expected to introduce a bias into the estimate of the GWB parameters \citep[see][]{2025A&A...694A..38F}.
As mentioned earlier, 
other four relevant sources of bias in EPTA \texttt{DR2new} data have been identified in \cite{goncharov2025readingsignaturessupermassivebinary}: (i) neglecting of epoch-correlated white noise;  (ii) neglecting the frequency dependence in the amplitude of the transient noise event in PSR J1713+0747; (iii) noise prior misspecification, which is solved through hierarchical inference \citep{Goncharov_2025,2024ApJS..273...23V}; and (iv) the exclusion of certain noise terms based on the results of single-pulsar noise analysis, which is solved by performing model averaging instead of model selection \citep{2025MNRAS.537L...1V}. Since epoch-correlated noise and transient noise in  PSR J1713+0747 were not injected in our simulations and the noise model used for the recovery is the same as the one used in the injection, effect (i), (ii) and (iv) are not affecting our results on the parameter estimation. This means that other effects are capable of generating a strong bias in \texttt{DR2new}. On the other hand, noise prior misspecification is expected to affect our recovery since we inject the noise components using values from the SPNA and then use uniform priors in the recovery. However, as will be shown in \autoref{sec:FL}, the main source of bias for the \texttt{DR2new} simulations appears to be the neglect of spectral leakage due to limited observation time in standard GWB analyses. See \cite{2025arXiv251014613Q} and \cite{2025arXiv250613866C} for in depth discussions on this effect.\\

Among the three selected realisations that resemble the real data also in the parameter estimation, realisation 94 seems to reproduce all the main features of the real data\footnote{Realisation 69 doesn't match so well the parameter estimation from the real data and realisation 70 has the maximum value of the HD S/N at 15yr, while real data peak at 10yr.} (\autoref{tab:specials}). The HD S/N distribution as a function of $T_{\rm obs}$ and the posterior distributions of the GWB parameters for realisation 94 are shown in \autoref{fig:94_snr} and \autoref{fig:94_pe}. We notice that this agreement is found despite the fact that parameter estimation depends heavily on the injected GWB spectrum, and we only tested one SMBHB population realisation.\\

From these two sections -- \ref{sec:FC} and \ref{sec:PE} --, we can conclude that the difference in the GWB significance and GWB parameter estimation found in the EPTA second data release when using the long (\texttt{DR2full}) and the short (\texttt{DR2new}) observation baselines is not an anomaly. Evidence of GWB compatible with the measured values is found in 5$\%$ of the simulations and we also found one realisation that agrees with the data in both the significance and the parameter estimation of the GW signal.
These percentages suggest that the observed scenario is consistent with an $\approx 2\sigma$ random fluctuations around the average behaviour.

\subsubsection{Spectral Leakage}\label{sec:FL}

\begin{figure*}[t]
   \centering
   \includegraphics[width=1\textwidth]{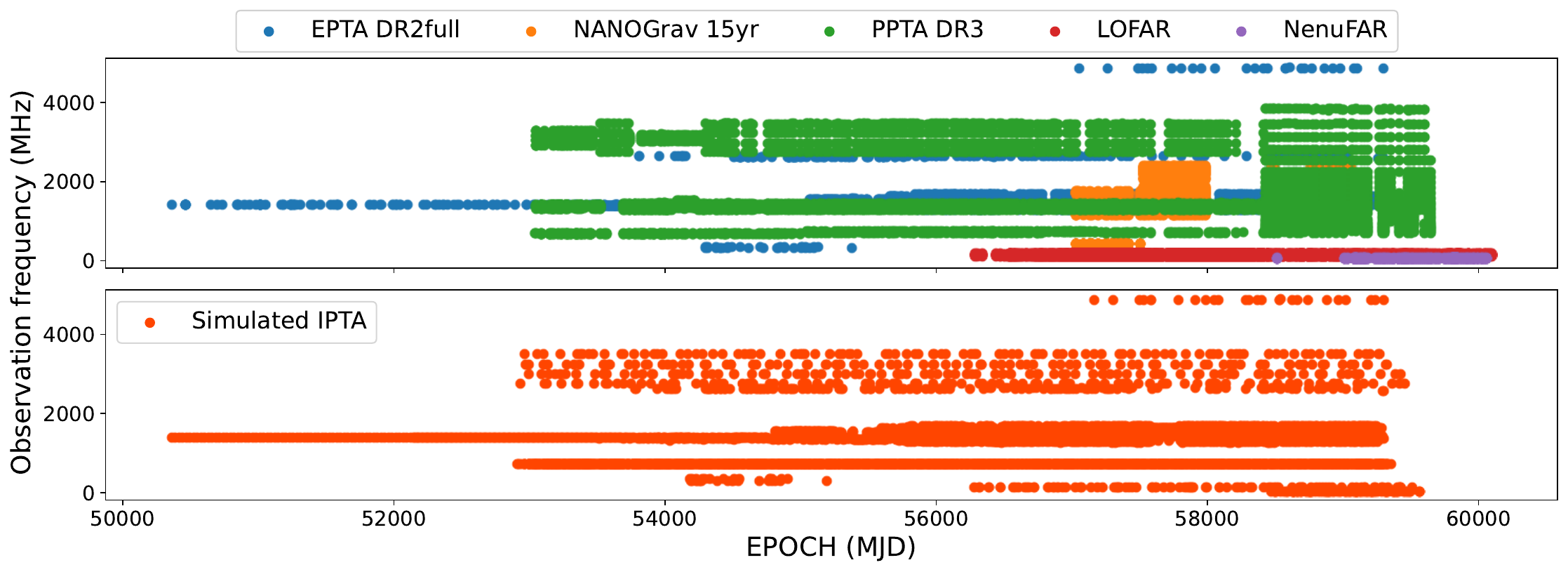}
   \caption{For pulsar PSR J1022+1001 {\rm Top panel}: data from the real EPTA \texttt{DR2full}, \texttt{NANOGrav 16yr}, \texttt{PPTA 20yr}, \texttt{LOFAR} and \texttt{NenuFAR}. {\it Bottom panel}: Simplified simulation of the three PTAs combination.}
   \label{fig:ipta_coverage}%
\end{figure*}

As mentioned in the previous section, we can identify three sources of the significant bias observed in the GWB parameter estimation from our simulations: the degeneracy between the individual noise and the GWB parameters, the effect of the spectral leakage and the noise prior misspecification. The first two effects are expected to be more prominent in \texttt{DR2new} than in \texttt{DR2full}.
In this section, we will focus on the spectral leakage, which has been proven to be the main source of bias in our \texttt{DR2new} simulations.\\
The fact that our data are observed on a finite time window introduces correlations between the Fourier frequency components that are neglected in the standard analysis, where the frequency-domain covariance matrix is considered diagonal. This approximation is valid for very long baseline observations, but can introduce a substantial bias in the parameter estimation when observation times are short. Since \texttt{DR2new} is only 10.3yr of observations, we expect the spectral leakage effect to be significant, and possibly the origin of the $\geq5\sigma$ bias in the GWB parameter estimation.
Let us point out that our \texttt{DR2new} simulations are expected to be affected by the spectral leakage in both the individual noise and in the signal covariance matrices. This is because (i) the noise is injected with the Fourier diagonal model, that is to say, on a harmonic Fourier basis that has a fundamental frequency of 1/$T_{\rm obs, psr}$ where $T_{\rm obs, psr} \in (13.6, 24.5){\rm yr}$ depending on the pulsar, and applying the 10.3 years rectangular window to obtain \texttt{DR2new} truncates the injected Fourier components to non-harmonic frequencies, (ii) the signal is injected as the sum of the exact time delays generated by every binary in the population with frequencies that do not match the 1/$T_{\rm obs, psr}$ harmonic Fourier basis.
To investigate this, we analyse the 100 realisations of \texttt{DR2new}  using the FFTInt method, as described in \cite{2025arXiv250613866C} and implemented in \texttt{Discovery}\footnote{\url{https://github.com/nanograv/discovery.git}}. Using the posterior distributions obtained from this analysis, we build the P-P plot with respect to $\gamma=13/3$, which is shown as the purple dashed line in \autoref{fig:PP_plot}. The significance of the bias goes from 5.6$\sigma$ with the standard analysis, to 1.5$\sigma$ with the FFTInt method. This means that in our \texttt{DR2new} simulations, among the three possible sources of bias in the GWB parameter estimation, the spectral leakage is the dominant one, while the other two effects only generate a bias that is within the precision of our recovery. 
Consequently, \texttt{DR2new} can be used to obtain a reliable estimate of the GW signal parameters, but only if the effect of the spectral leakage is taken into account.

It is worth noticing that by extending the analysis to frequencies lower than $1/T_{\rm obs}$ -- in particular, a cutoff at $1/3T_{\rm obs}$ was used -- the estimate of the GWB parameters slightly changes: the difference in the estimate of $\gamma_{\rm GWB}$ is $\Delta\gamma_{\rm GWB}\sim-0.12\pm0.14$ quoting the 68$\%$C.I. This means that extending the frequency range gives, on average, a shallower GWB recovery, with a difference that is compatible with zero.

\begin{figure*}[!t]
    \centering
    \begin{subfigure}[t]{0.48\textwidth}
        \centering
        \includegraphics[width=\linewidth]{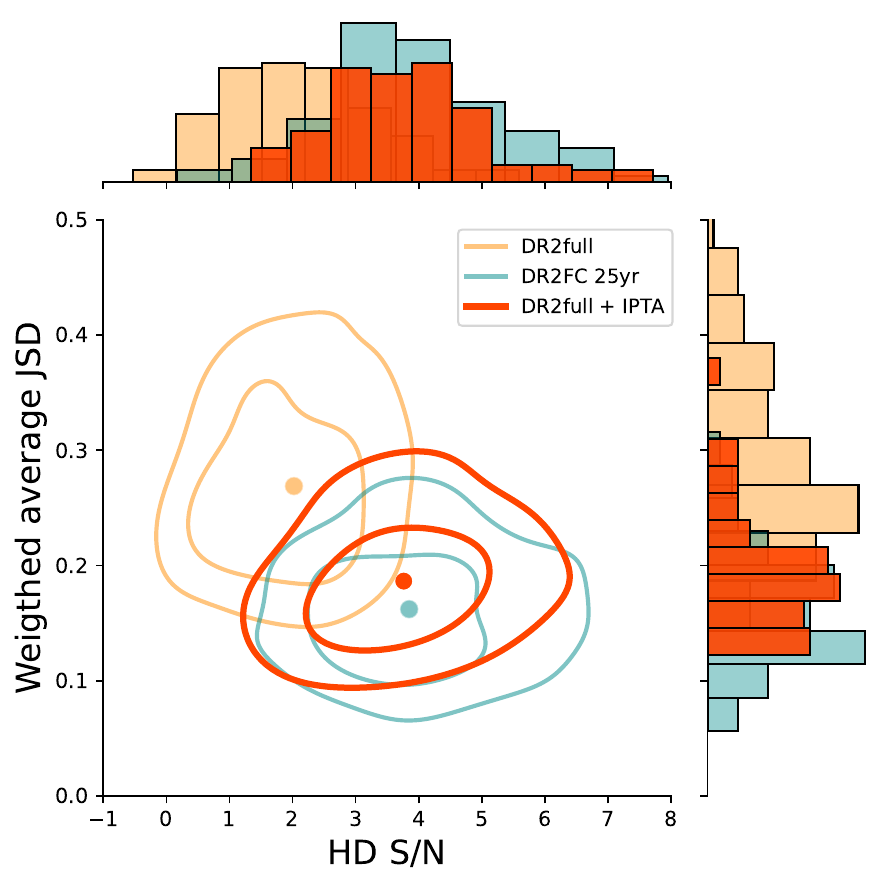}
        \caption{2-d distributions of the noise marginalised HD S/N and the degeneracy coefficient (weighted average JSD).}
        \label{fig:SNR_JSD_ipta}
    \end{subfigure}
    \hfill
    \begin{subfigure}[t]{0.48\textwidth}
        \centering
        \includegraphics[width=\linewidth]{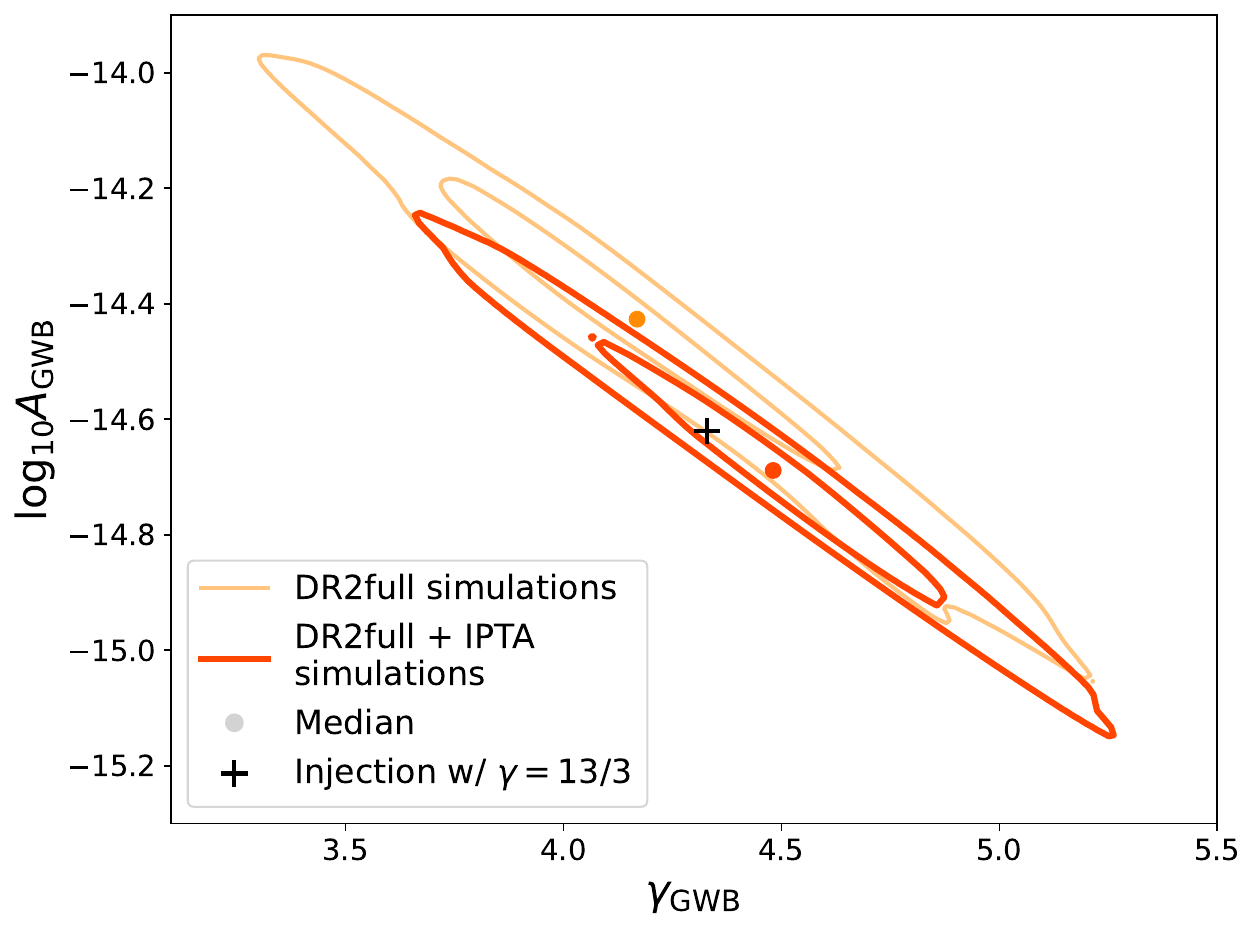}
        \caption{Distribution of the median of log$_{10}A_{\rm GWB}$ and $\gamma_{\rm GWB}$ marginalised posterior distributions.}
        \label{fig:PE_ipta}
    \end{subfigure}

    \begin{subfigure}[t]{0.95\textwidth}
        \centering
        \includegraphics[width=\linewidth]{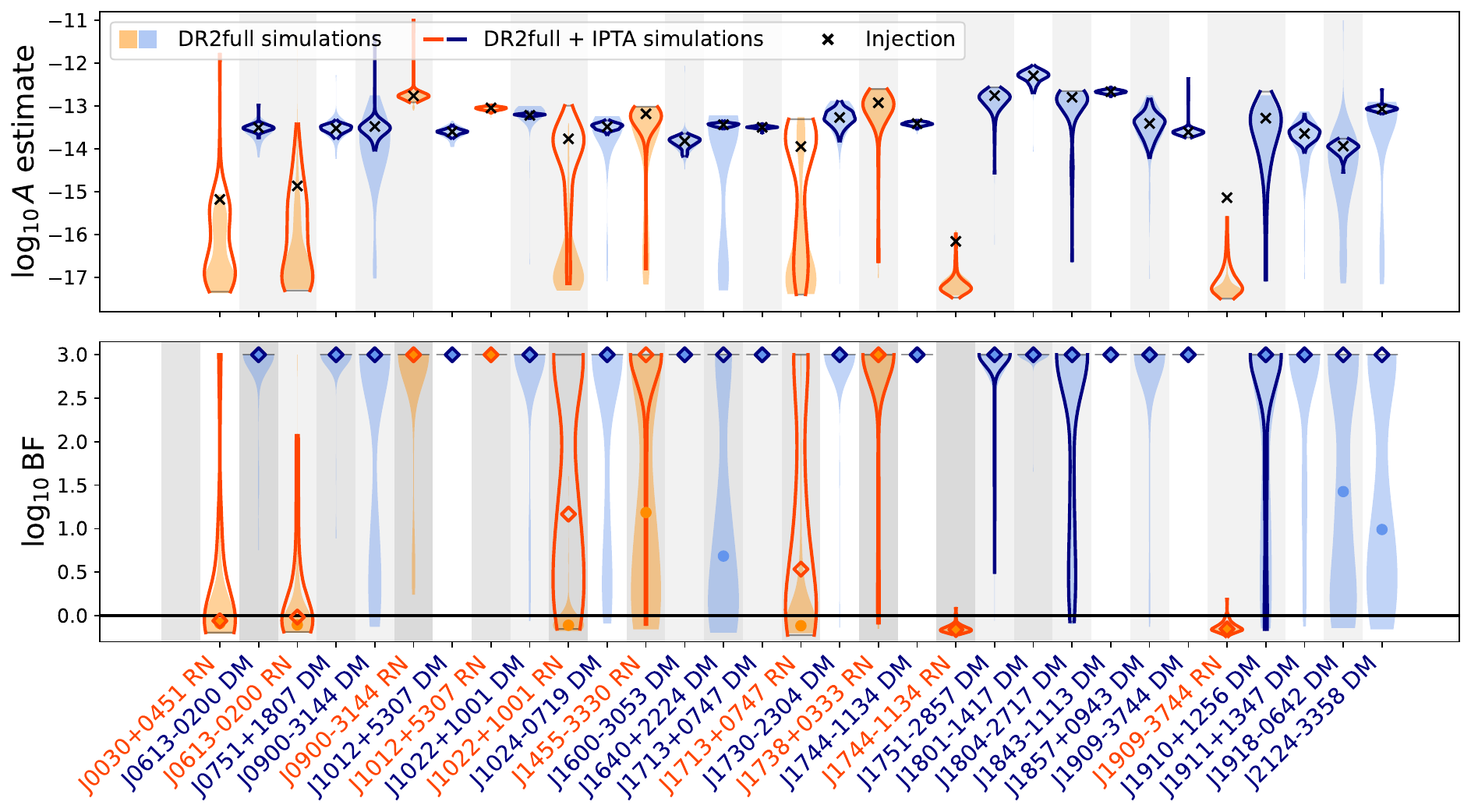}
        \caption{
        Summary of the noise parameters estimation from the 100 realisations of \texttt{DR2full} (orange) and \texttt{DR2full + IPTA} (red) simulated data. {\it Top panel}: median of the marginalised posterior distribution of the noise component amplitude, the true values are shown for reference (black crosses). {\it Bottom panel}: Savage-Dickey ratio of the amplitude posterior distributions; BF < 1 means the posterior is unconstrained, BF $\gg 1$ means the posterior is very well constrained.}
        \label{fig:noise_pe}
    \end{subfigure}

    \caption{GWB significance and parameter estimation from the 100 \texttt{DR2full + IPTA} simulations.}
    \label{fig:combined_figure}
\end{figure*}

\subsection{IPTA frequency coverage}\label{sec:ipta}

In Section~\ref{sec:FC}, we found that the limited frequency coverage in the first half of \texttt{DR2full} significantly restricts its performance in the GWB search. However, there are several feasible ways to increase the frequency coverage of our data, and these are currently being implemented. Indeed, the next data release of the International Pulsar Timing Array \citep[IPTA, e.g.][]{2016MNRAS.458.1267V} will combine data from all the main PTA collaborations: EPTA, InPTA, NANOGrav, PPTA, MPTA and CPTA plus data from the LOw Frequency ARray \citep[LOFAR,][]{2020A&A...644A.153D} and from the New extension in Nan\c{c}ay upgrading \citep[NenuFAR,][]{2020A&A...644A.153D}. This will allow to build an array of more than 100 pulsars, with many of the pulsars observed by multiple PTAs in different frequency bands. In particular, LOFAR and NenuFAR, NANOGrav and PPTA are very useful to complement the first problematic 10 years of the EPTA \texttt{DR2}. The reason is that NANOGrav and PPTA, with observation times of respectively $\sim 16$yr and $\sim20$yr, have long baseline measurements at frequencies complementary to EPTA; LOFAR and NenuFAR, despite having a shorter baseline ($\sim$11yr) collects data at extremely low frequencies (100-200MHz for LOFAR and 30-90MHz for NenuFAR) for 12 EPTA pulsars. For more details on the LOFAR and NenuFAR datasets and the advantages of their combination with the EPTA \texttt{DR2}, refer to \cite{2025arXiv251004639I}. The aim of this section is to simulate in a simplified, yet representative, way how the 25 pulsars of EPTA would behave once these 4 complementary datasets are added. Therefore, we added to the \texttt{DR2full} simulations the following data:
\begin{itemize}
    \item four high-frequency backends (at ~2750, 3000, 3250 and 3500 MHz) starting from EPOCH = 53000 with TOA error of 2$\mu$s, which resemble PPTA-like backends
    \item one low-frequency backend (750 MHz) starting from EPOCH = 53000 with TOA error of 2$\mu$s, which in the real data is given either by NANOGrav or by PPTA
    \item 100 TOAs at 150MHz, with TOA error of 7$\mu$s, which are typical properties of LOFAR data, starting from EPOCH = 56500
    \item 100 TOAs at 60MHZ starting from EPOCH = 58300 and with error of 7$\mu$s, which are typical of NenuFAR
\end{itemize}
These backends are only added to those pulsar for which they are indeed available in the real data. This method allows us to have a representative simulation of how EPTA pulsars will look like in the next IPTA data release. MPTA and CPTA backends, despite being the most precise TOAs in the IPTA, are not considered here because they have very short baselines.

An example of the frequency coverage of this simulated dataset is shown in the bottom panel of \autoref{fig:ipta_coverage}. In the top panel, the real TOAs from the 5 PTA involved are shown. This analysis aims to assess whether the additional frequency coverage provided by the IPTA collaboration is sufficient to solve the degeneracy problem affecting \texttt{DR2full}. This version of \texttt{DR2full} will be referred to as \texttt{DR2full + IPTA}.\\

Figure~\ref{fig:SNR_JSD_ipta} shows the joint distribution of the degeneracy coefficient and the noise-marginalised HD S/N for 100 realisations of \texttt{DR2full + IPTA}, compared with \texttt{DR2full} and \texttt{DR2FC}. Both the parameter degeneracy and the HD S/N improve significantly with IPTA coverage. The HD S/N median almost doubles, rising from $2.0^{+1.3}_{-1.0}$ in \texttt{DR2full} to $3.8^{+1.2}_{-1.0}$ in \texttt{DR2full + IPTA}. The performance of this dataset is equal to the performance obtained with the homogeneous frequency coverage of \texttt{DR2FC}. This is true also when looking at the evolution of the HD S/N with the observation time (not shown), which is equivalent to the one of  \texttt{DR2FC} displayed in \autoref{fig:SNRTobs}: the HD S/N keeps growing with increasing $T_{\rm obs}$ and there are no realisations in which the 10yr version of the dataset gives a higher significance than the 25yr one. In the five ``pathological'' realisations resembling the data, the HD S/N always exceeds 3 when IPTA backends are added, compared to a maximum of 2.2 in \texttt{DR2full} (\autoref{tab:specials}).\\

Parameter estimation also benefits from the improved coverage: both the amplitude and slope of the GWB are accurately recovered (\autoref{fig:PE_ipta}), with the slope bias reduced to a non-significant $0.8\sigma$ ($\sim 3\sigma$ from \texttt{DR2full}). Moreover, the overall precision on GWB parameters improves by $\sim25\%$ in amplitude and $\sim40\%$ in slope.\\

Finally, it is interesting to observe how the addition of these complementary backends affects the estimation of the noise parameters. \autoref{fig:noise_pe} shows the distribution of the posterior distribution median for the amplitude of each noise process involved in the search, compared to the injected value (top panel), and the constraining power on each parameter (bottom panel). The constraining power is quantified by the Bayes factor, which is approximated by the Savage–Dickey ratio: if the ratio is smaller than 1, then the posterior distribution is not constrained. When the ratio is at its maximum value (set to 1000 for representation purposes), the parameter is very precisely constrained.
Several observations can be made from this plot; most importantly, all parameters that the model can constrain are centred on the correct injected value, indicating unbiased noise parameter estimation.
Now focusing on the impact of IPTA coverage on the estimation of the noise parameters (see \autoref{fig:noise_pe}), we can see that only in 15 pulsars out of 21 the DM parameters are constrained in all realisations of \texttt{DR2full}, whereas, when the IPTA backends are added, all the DM parameters are very well constrained in every realisation. This allows to mitigate degeneracy of the these parameters with the GWB.
Conversely, RN parameters are typically unconstrained or poorly constrained in \texttt{DR2full}, except when the amplitude is very high ($\gtrsim 10^{-13}$), as it is in PSR J0900-3144 and PSR J1012+5307. And despite the addition of IPTA frequency coverage, constraints on these parameters remain poor. However, a significant improvement can be seen in the cases of PSR J1022+1001, J1455-3330 and J1713+0747. These are the only three pulsars with an amplitude between $10^{-14}$ and $10^{-13}$. This suggests that \texttt{DR2full + IPTA} frequency coverage is sufficient, in many realisations, to constrain red noise processes with an amplitude greater than 10$^{-14}$. It is therefore reasonable to assume that even better frequency coverage and higher S/N would allow us to constrain all red noise parameters, also solving their degeneracy with the GWB.\\

We can therefore conclude from this section that adding the longest baseline backends of NANOGrav and PPTA and the low frequency backends of LOFAR and Nenufar to EPTA \texttt{DR2full} is sufficient to significantly improve the evidence in favour of the GW signal, as well as to provide an accurate and precise measurement of the GWB parameters. This conclusion supports the findings of \cite{2025MNRAS.542.3028L}, which demonstrate that partial IPTA-style data combinations, such as the Lite method, can rapidly improve PTA sensitivity by selecting the most informative pulsar datasets prior to full formal combination.
An interesting obervation is that the addition of NANOGrav and PPTA boosts the average HD S/N from 2.0 to $\sim 3.2$. The further addition of LOFAR and NenuFAR backends boosts it from $\sim3.2$ to 3.8, indicating that adding very low frequency backends has an effect that is comparable to adding complementary long baseline observations.

\subsection{The 5yr dataset - impact of the time span on the parameter estimation}\label{sec:5yr}

In this last section, we investigate the differences between long- and short-timespan datasets, highlighting how the estimation of the GWB parameters depends on the observation time. To do this, we analyse the longest and shortest simulated datasets produced for this work: the 25yr and 5yr datasets, referred to as \texttt{DR2full} and \texttt{DR2 5yr}.

\begin{figure*}[!t]
    \centering
    \begin{subfigure}{0.48\textwidth}
        \centering
        \includegraphics[width=\linewidth]{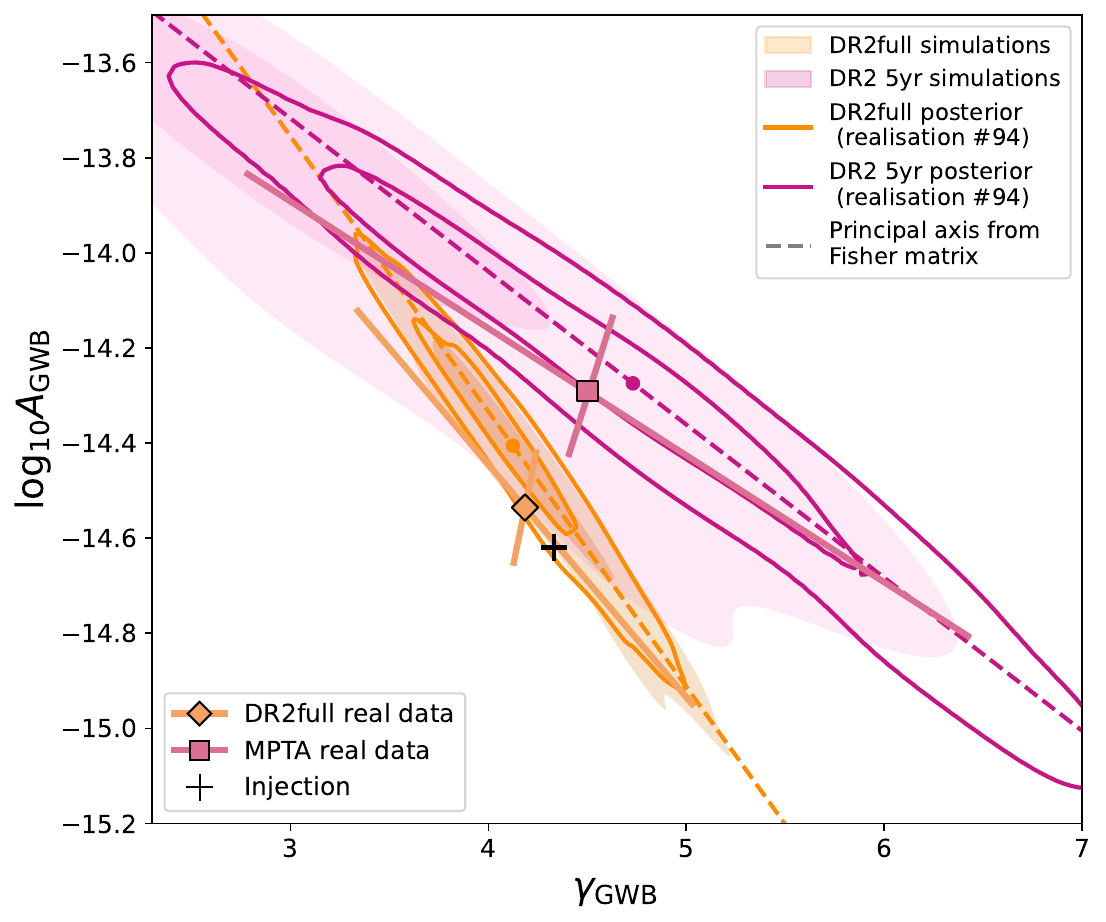}
        \caption{Comparison between \texttt{DR2full} and \texttt{DR2 5yr}. For comparison, the principal axes of the real EPTA \texttt{DR2}full and MeerKAT PTA are also shown.}
        \label{fig:PE_Fisher}
    \end{subfigure}
    \hfill
    \begin{subfigure}{0.48\textwidth}
        \centering
        \includegraphics[width=\linewidth]{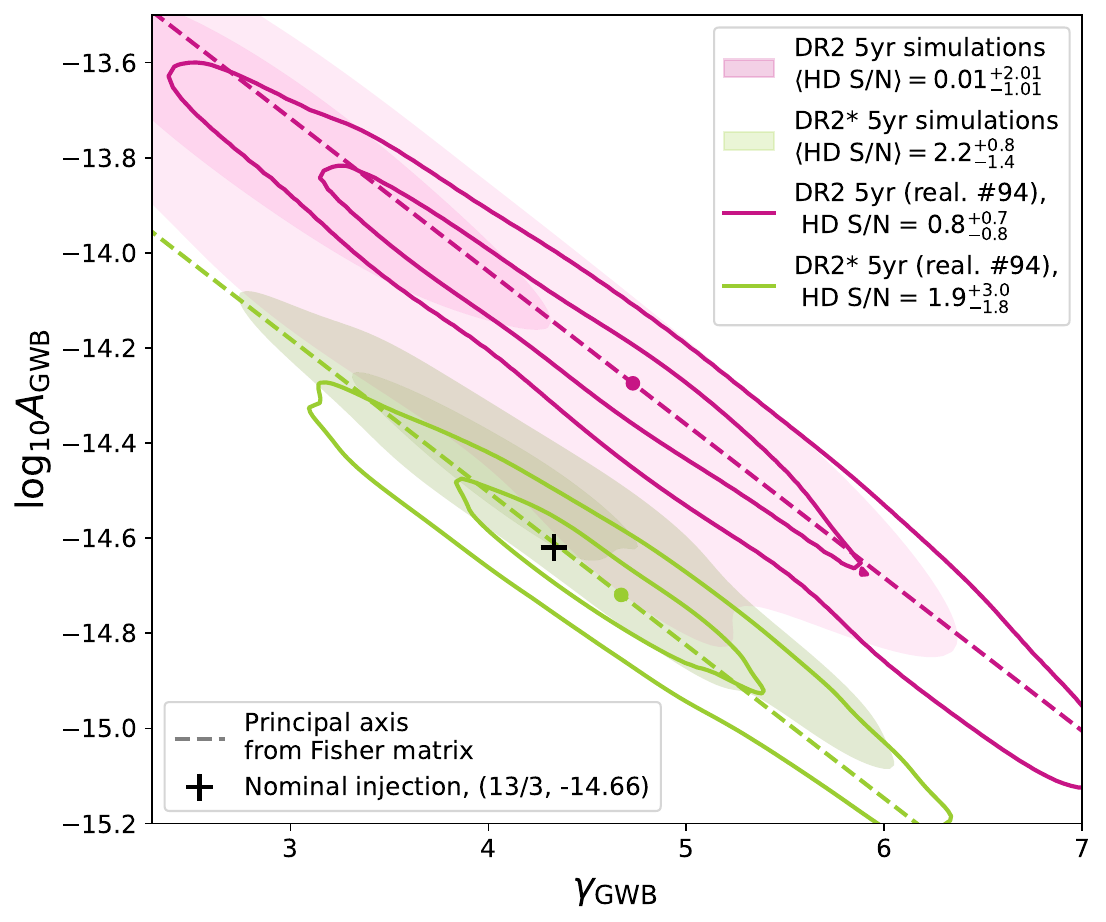}
        \caption{Comparison between \texttt{DR2* 5yr} and \texttt{DR2 5yr}. For reference, the injected value is displayed as a black cross.}
        \label{fig:PE5yr_Fisher}
    \end{subfigure}
    \caption{Median of the marginalised posterior distributions of $\gamma_{\rm GWB}$ and $\log_{10}A_{\rm GWB}$ of all the 100 simulations ({\it shaded region}), posterior distribution of realisation 94 ({\it solid line}) and principal axis evaluated orientation evaluated from the Fisher matrix ({\it dashed line}).}
    \label{fig:PE_tspan}
\end{figure*}

In \autoref{fig:PE_tspan}, we show the distribution of the parameter estimates obtained from the 100 realisations of both datasets, together with the posterior distribution of a single realisation. The chosen realisation is number 94, selected in \autoref{sec:PE} as the most similar to the real EPTA \texttt{\texttt{DR2}}.
The first observation is that the posterior obtained from the 5yr dataset is much broader, as expected due to the shorter timespan. However, its principal axis also exhibits a different orientation compared to that of the 25yr dataset. Interestingly, a similar behaviour is observed in real datasets: in the same figure, we display the principal axes of the posterior distributions obtained from EPTA \texttt{DR2full} (24.8 yr) and from the \texttt{MeerKAT PTA} (4.5 yr). These two posteriors show a significant difference in orientation, consistent with the results from our simulations. 

This dependence of the posterior orientation on the observation timespan arises from the functional form of the likelihood and can be predicted using the Fisher formalism. Following the derivation in \cite{2024PhRvD.110f3022B}, the orientation of the principal axis of the posterior can be computed analytically and is shown as a dashed line in \autoref{fig:PE_Fisher}. The agreement between the Fisher prediction and the simulation results is very good. Indeed, by computing the orientation for all the 100 noise realisations, we found that the prediction from the Fisher matrix agrees with the distribution of the orientation values within 68$\%$C.I. (see \autoref{appendix:B}) for all the observation times investigated. In addition, the values of the orientation computed from the posterior obtained with the real EPTA \texttt{DR2full}, \texttt{DR2new} and \texttt{MPTA} are compatible with the 90$\%$C.I. of the simulations.

\begin{figure}[t]
   \centering
   \includegraphics[width=0.5\textwidth]{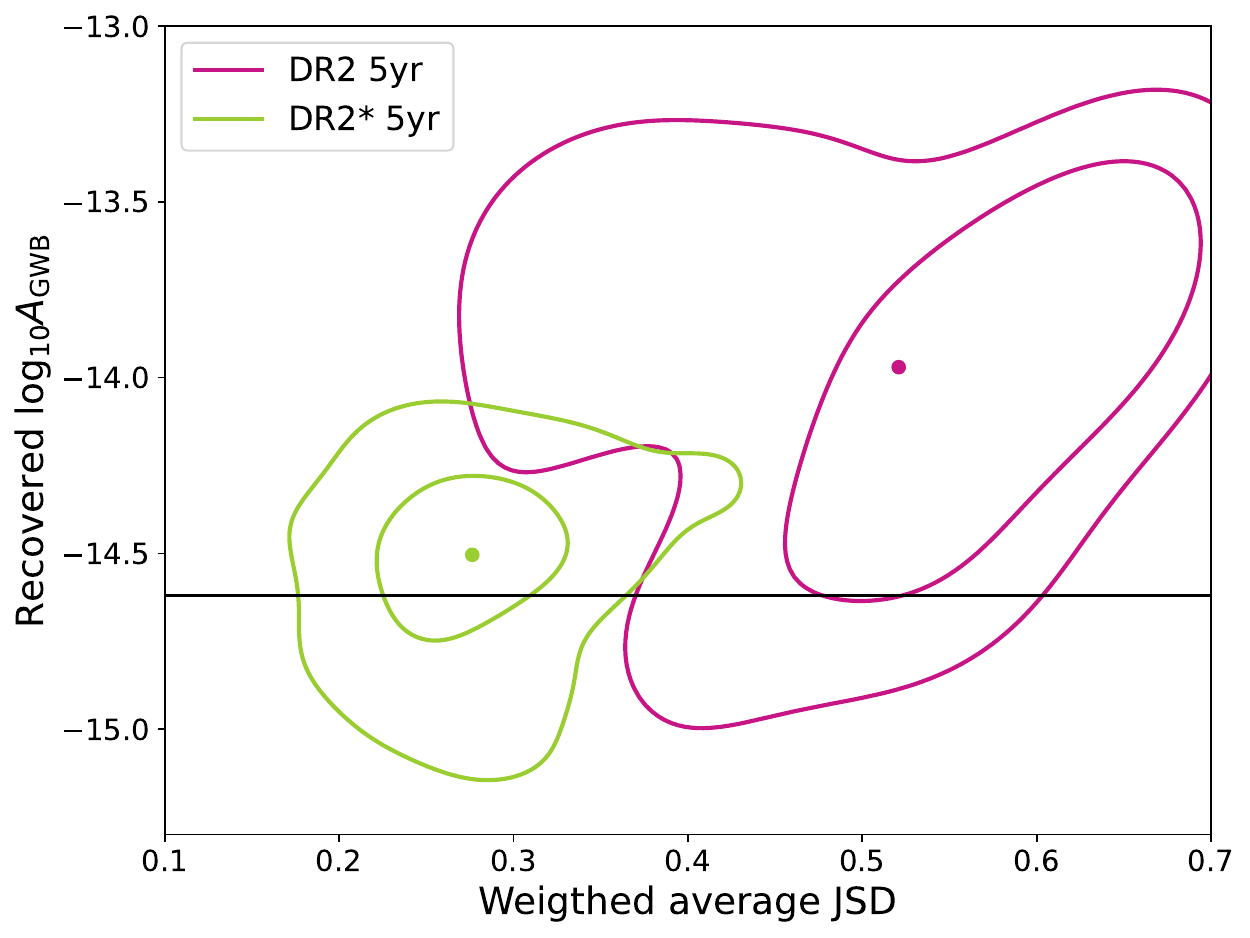}
   \caption{Degeneracy coefficient (weighted averag JSD) and median of the marginal posterior distribution of log$_{10}A_{\rm GWB}$ from the 100 simulations of \texttt{DR2 5yr} and \texttt{DR2* 5yr}.}
   \label{fig:5yr_pejsd}%
\end{figure}

The increase in the principal axis angle over time can be explained by the fact that the uncertainty in the GWB slope is more sensitive to the observation time than the uncertainty in the amplitude. Indeed, by denoting the variances of the amplitude and slope as $\sigma_A^2$ and $\sigma_\gamma^2$ respectively, an analytical calculation yields the following proportionality:
\begin{equation}\label{eq:sT}
    \begin{cases}
    \sigma_{A}^2 \propto ({\rm S/N})^{-2}\\
    \displaystyle \sigma_{\gamma}^2 \propto ({\rm S/N})^{-2}\left[\ln\left(\frac{{\rm yr}}{T_{\rm obs}}\right)\right]^{-2}
    \end{cases}
\end{equation}
where the additional dependence of $\sigma_{\gamma}^2$ on $T_{\rm obs}$ is explicit. For the derivation and a detailed discussion, see \autoref{appendix:B}.

Equation~\ref{eq:sT} also shows that the orientation of the posterior distribution does not depend on the overall ${\rm S/N}$, since both variances scale equally with it. To test this, we simulated a high-${\rm S/N}$ version of the 5yr dataset, that we will refer to as \texttt{DR2* 5yr}. The average S/N increases from $\sim 0.01$ (for \texttt{DR2 5yr}) to $\sim 1.9$ (for \texttt{DR2* 5yr}). The distribution of medians and a representative posterior for this high-${\rm S/N}$ dataset are shown in \autoref{fig:PE5yr_Fisher}, where we can see that the posterior orientation remains roughly the same when increasing the S/N, consistent with the Fisher prediction.

An interesting implication of this timespan dependence can be seen by comparing two datasets with similar HD ${\rm S/N}$ but very different observation times: this is the case for our simulated \texttt{DR2full} (25\,yr, average HD ${\rm S/N}\approx2$) and \texttt{DR2* 5yr} (5\,yr, average HD ${\rm S/N}\approx1.9$). Although the two datasets provide comparable precision on the GWB amplitude (68\% C.I. of $0.38$ and $0.45$, respectively), they yield significantly different constraints on the GWB slope, with a 68\% C.I. of $0.71$ for \texttt{DR2full} and of $1.6$ for \texttt{DR2* 5yr}.\\

The posterior distributions obtained from the short-timespan dataset are not only less precise than those obtained from the long-timespan dataset, but also less accurate, as it can be noticed from \autoref{fig:PE5yr_Fisher}. This is to be expected; we have already seen an analogous situation when comparing the parameter estimation obtained from \texttt{DR2full} and \texttt{DR2new} (see \autoref{sec:PE}). However, while the majority of the bias in the case of \texttt{DR2new} was due to neglecting the inter-frequency correlation and was solved by taking this spectral leakage into account in the model (see \autoref{sec:FL}), something slightly different is occurring in \texttt{DR2 5yr}. By building the P-P plot, we find that the \texttt{DR2 5yr} dataset has a bias in the parameter recovery with a significance of approximately 6.4$\sigma$, which reduces to $\sim3.8\sigma$ if we account for spectral leakage in the analysis. In contrast, applying the standard diagonal analysis to \texttt{DR2* 5yr} — a version of \texttt{DR2 5yr} with high HD S/N — renders the bias in the recovery insignificant ($\sim0.95\sigma$). The most straightforward explanation for these results is that most of the bias is due to the degeneracy of the pulsar individual noise with the GWB parameters, since the HD S/N is so low and therefore it is hard to tell individual noise apart from the common process. Indeed, as can be seen in \autoref{fig:5yr_pejsd}, the high and low S/N datasets occupy two different regions in the {\it degeneracy coefficient – recovered amplitude} space, and there is a strong correlation between the two quantities ($r\sim0.5$, with significance greater than 5$\sigma$). Thus, the higher amplitude and lower spectral index in the recovery from \texttt{DR2 5yr} are related to the strong degeneracy of the GWB parameters with the individual noise parameters. For \texttt{DR2new}, the degeneracy was much smaller thanks to the higher HD S/N. On the other hand, the spectral leakage is expected to affect more \texttt{DR2 5yr} than \texttt{DR2new} because of the shorter data span, which is not what we observe. This can be explained by the effect of timing model marginalisation, which acts as a windowing function and mitigates the effects of spectral leakage \citep[see][]{2025arXiv250613866C, 2025arXiv251014613Q}. In \texttt{DR2 5yr}, the GWB spectrum is modelled with only 4 components and the fit is likely dominated by the first one or two bins, which are the most affected by the TM marginalisation. Consequently the timing model marginalisation manages to mitigate the effect of the spectral leakage. This mitigation is less effective in \texttt{DR2new} because more frequency components contribute to the GWB parameter estimation in that case.

\section{Conclusions}
\label{sec:con}

As Pulsar Timing Array experiments enters the detection era, it is crucial that we develop a comprehensive understanding of our data and of the outcomes of our analysis pipelines. This work is a part of this effort and it focuses, in particular, on the last data release of the EPTA collaboration. This dataset was presented in two versions, EPTA \texttt{DR2full}, which has an observation timespan of 24.8yr, and \texttt{DR2new}, that includes only the last 10.3yr of \texttt{DR2full} \citep{wm3}. The unexpected result obtained by the analysis of these two datasets was that the shorter version gave a higher significance in favour of the presence of a GWB (HD S/N $\sim3.5$) than the longer version (HD S/N $\sim1.3$). This counterintuitive result motivated the analysis carried out in this paper. A faithful set of \texttt{DR2full}-like simulations has been generated and then cut at different epochs (20, 15, 10 and 5 years before the most recent TOA) to obtained shorter versions of the dataset, among which \texttt{DR2new}-equivalent simulations (10yr dataspan). For each time span, 100 different noise realisations are injected, on top of the same stochastic GW signal. The outcome of the analyses run on all these simulations led us to the following conclusions:
\begin{itemize}
    \item the first 10 years of \texttt{DR2} are not sufficiently informative to contribute significantly to the GWB detection, mainly due to the limited frequency coverage in this part of the data. As a result, the HD S/N distributions for \texttt{DR2new} and \texttt{DR2full} largely overlap. This implies that, due to random fluctuations in the HD S/N, \texttt{DR2new} can occasionally yield a higher GWB significance than \texttt{DR2full}, which occurs in about 15$\%$ of the realisations. In fact, the probability of observing a scenario similar to that seen in the EPTA \texttt{DR2} within our controlled simulations is 5$\%$. In these cases, adding the first 10 or 15 years of data degrades the signal, instead of enforcing it. Moreover, a realisation was found that reproduced the data behaviour exactly, both in terms of GWB significance and parameter estimation. This shows that the behaviour observed in \texttt{DR2new} vs \texttt{DR2full} could well be simply due to an $\approx 2\sigma$ occurrence of noise fluctuations, with no need to invoke other problems or issues with the data or the analysis pipelines.
    \item \texttt{DR2new}-like simulations lead to a significantly biased GWB parameter estimation, with bias being towards higher values of the amplitude and lower values of the slope. This bias is related to the correlation between different frequency bins, which is neglected in the standard analysis. If spectral leakage (i.e. frequency bin correlation) is accounted for in the model -- by using the method proposed in \cite{2025arXiv250613866C} -- \texttt{DR2new} also provides a reliable estimate of the GWB parameters. Therefore, this analysis emphasises the importance of accounting for spectral leakage in GWB analyses.
    \item Adding long-baseline complementary backends (such as TOAs from NANOGrav and PPTA) and low-frequency measures (such as those obtained with LOFAR and NenuFAR) to \texttt{DR2full} is crucial to address its pathological behaviour. Indeed, when these backends are added — in a simplified, yet realistic simulation — the average HD S/N increases from 2.0 in \texttt{DR2full} to 3.8 in \texttt{DR2full + IPTA}, and the HD S/N trend with observation time is monotonically increasing. This proves that the first 10 years of data significantly contribute to the HD S/N, and \texttt{DR2full} has the potential of prividing high S/N and precise and accurate parameter estimation when combined with other datasets within the IPTA framework.
    \item Very short baseline datasets with a low HD S/N ratio (below 2) are greatly impacted by spectral leakage and the degeneracy between individual noises and the GWB parameters. This results in a significant bias in the estimation of GWB parameters towards high amplitudes. Therefore, modelling the noise and accounting for spectral leakage in the analysis is crucial to obtain a reliable result from such short datasets. This could be particularly relevant when analysing the first releases of CPTA and MeerKAT data. Moreover, studying these short-baseline datasets allows us to investigate how the precision of the GWB parameter estimates evolves with observation time $T_{\rm obs}$. $T_{\rm obs}$ has a greater impact on the precision of $\gamma$ than on log$_{10}A$, which means that long-baseline datasets are important for maximising our ability to constrain the GWB slope, which carries valuable information about the eccentricity and environmental coupling of the underlying SMBHB population.
\end{itemize}

These results demonstrate that the behaviour observed in EPTA \texttt{DR2} is consistent with the outcome of current analysis pipelines on controlled simulated data, without necessarily indicating any suspicious anomalies. They also highlight the importance of accurate GWB modelling along with broadband frequency coverage and joint PTA analyses for robust GWB detection.

\begin{acknowledgements}
IF would like to acknowledge the precious help and useful discussions of Marco Crisostomi, Hippolyte Quelquejay Leclère and Aurélien Chalumeau, which contributed to shaping this paper. AS acknowledges the financial support provided under the European Union’s H2020 ERC Advanced Grant ``PINGU'' (Grant Agreement: 101142079).
\end{acknowledgements}

\appendix

\section{Special realisation}
\begin{figure*}[!ht]
    \centering
    \begin{subfigure}{0.48\textwidth}
        \centering
        \includegraphics[width=\linewidth]{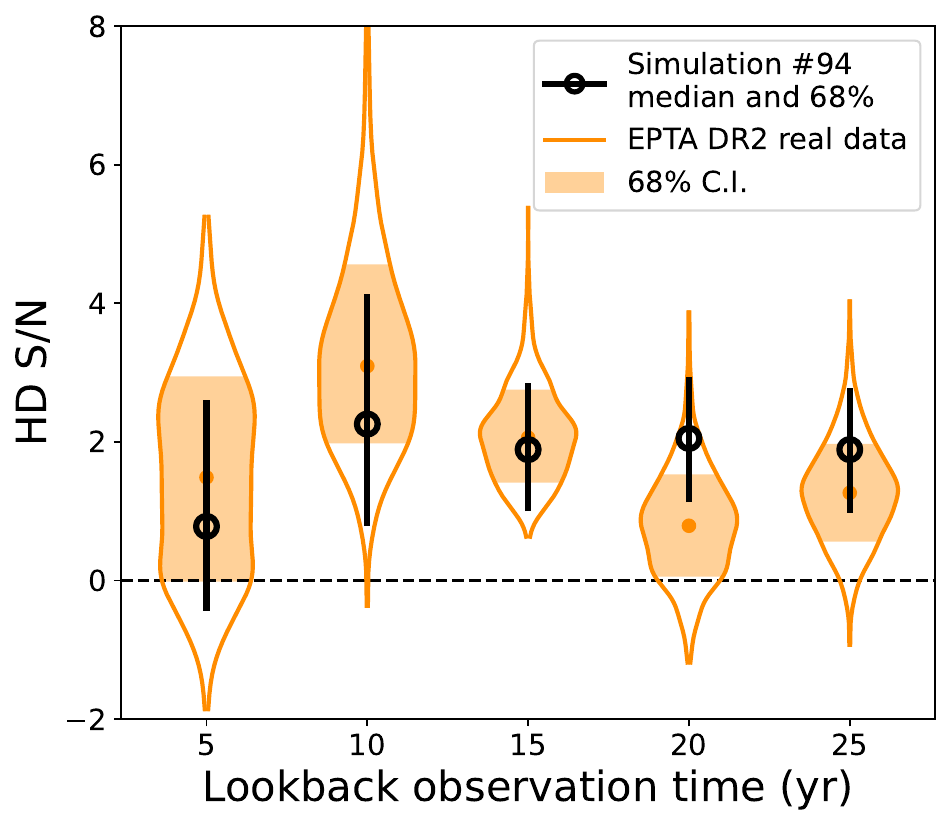}
        \caption{ }
        \label{fig:94_snr}
    \end{subfigure}
    \hfill
    \begin{subfigure}{0.48\textwidth}
        \centering
        \includegraphics[width=\linewidth]{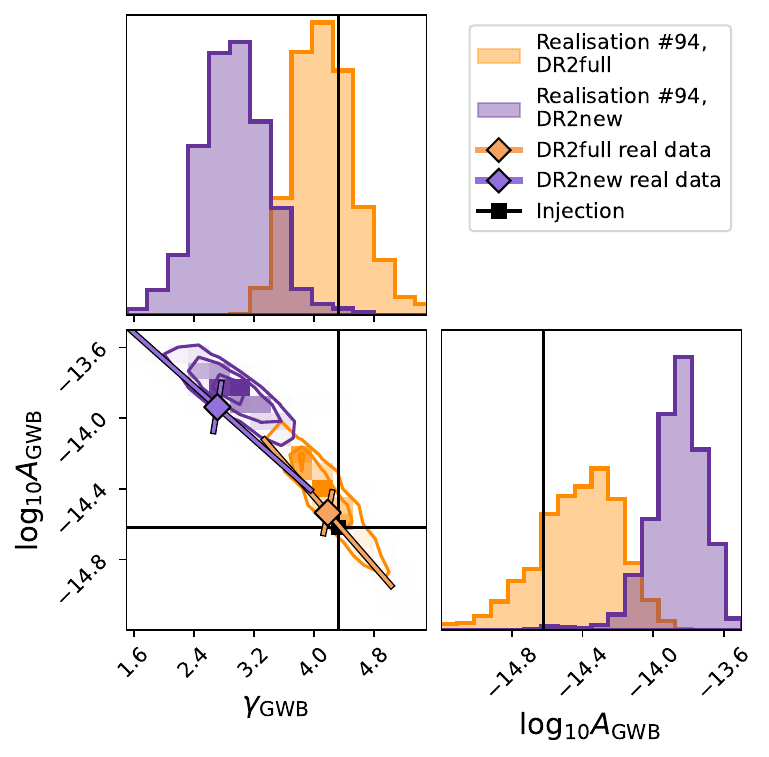}
        \caption{ }
        \label{fig:94_pe}
    \end{subfigure}
    \caption{(a) Median and 68$\%$C.I. of the noise marginalised HD S/N computed with realisation 94 by cutting the dataset at different epochs. The same cuts were applied to the real EPTA \texttt{DR2} and the violins shown the HD S/N distribution. (b) Marginalised posterior distributions of $\gamma_{\rm GWB}$ and $\log_{10}A_{\rm GWB}$ obtained from the \texttt{DR2full} and \texttt{DR2new} version of realisation 94. As a reference, the median values and 2$\sigma$ covariance principal axes obtained from the real data are displayed.} 
    \label{fig:94}
\end{figure*}

Among the 100 realisations of noise that were generated, 5 reproduce well the behaviour of the real data in the HD S/N evolution with time. Among these 5 realisations, one behaves very similar to the data both in the HD S/N and in the parameter estimation. In figure \autoref{fig:94_snr} the evolution of the HD S/N with the observation time is shown for both the realisation 94 and the real data. In \autoref{fig:94_pe} the parameter estimations obtained from the 25yr and the 10yr versions of realisation 94 and the real data are compared. It can be seen ho the simulated and the real data agree well with one another.

\clearpage
\section{Fisher matrix calculation}\label{appendix:B}
The expression of the Fisher matrix according to \cite{2024PhRvD.110f3022B} is
\begin{equation}
\begin{aligned}
F_{\alpha\beta} = & \sum_{f_k}C_{IJ}^{-1}C_{KL}^{-1}\frac{\partial (R_{JK}S_{\rm h}(f_k))}{\partial \theta^{\alpha}}\frac{\partial (R_{IL}S_{\rm h}(f_k))}{\partial \theta^{\beta}} =\\
& = \sum_{f_k}C_{IJ}^{-1}C_{KL}^{-1}R_{JK}R_{IL}\frac{\partial S_{\rm h}(f_k)}{\partial \theta^{\alpha}}\frac{\partial S_{\rm h}(f_k)}{\partial \theta^{\beta}} =\\
& = \sum_{f_k} \frac{T_{\rm obs}}{S_{\rm eff}^2}\frac{\partial S_{\rm h}}{\partial \theta^{\alpha}}\frac{\partial S_{\rm h}}{\partial \theta^{\beta}}
\end{aligned}
\end{equation}
where 
\begin{equation}
S_{\rm eff} = \left[\frac{1}{T_{\rm obs}}\cdot C_{IJ}^{-1}C_{KL}^{-1}R_{JK}R_{IL}\right]^{-1/2}
\end{equation}
and 
\begin{equation}
{\rm S/N}^2 = T_{\rm obs}\sum_{f_k}\left(\frac{S_{\rm h}}{S_{\rm eff}}\right)^2
\end{equation}
The partial derivatives take the simple form
\begin{equation}
\begin{cases}
\displaystyle \frac{\partial S(f)}{\partial {\rm logA}} = 2{\rm ln10} \cdot S(f)\\
\displaystyle \frac{\partial S(f)}{\partial {\rm \gamma}} = -{\rm ln}\left(\frac{f}{\rm yr^{-1}}\right) \cdot S(f)
\end{cases}
\end{equation}
thus the entrances of the covariance matrix are
\begin{equation}
\begin{cases}
\displaystyle F_{AA} = (2{\rm ln10})^2\cdot T_{\rm obs}\sum_{f_k}\left(\frac{S_{\rm h}}{S_{\rm eff}}\right)^2 = (2{\rm ln10})^2 \cdot {\rm S/N}^2\\
\begin{aligned}
        \displaystyle F_{\gamma \gamma} & = T_{\rm obs}\sum_{f_k}\left[-{\rm ln}\left(\frac{f_k}{\rm yr^{-1}}\right) \right]^2\left(\frac{S_{\rm h}}{S_{\rm eff}}\right)^2 \approx\\
& \approx\left[{\rm ln}\left(\frac{f_{\rm min}}{\rm yr^{-1}}\right)\right]^2\cdot {\rm S/N}^2
\propto \left[{\rm ln}\left(\frac{\rm yr}{T_{\rm obs}}\right)\right]^2\cdot {\rm S/N}^2
\end{aligned}\\
\begin{aligned}
\displaystyle F_{A \gamma} & = 2{\rm ln10}\cdot T_{\rm obs}\sum_{f_k}\left[-{\rm ln}\left(\frac{f_k}{\rm yr^{-1}}\right) \right]\left(\frac{S_{\rm h}}{S_{\rm eff}}\right)^2 \approx \\
& \approx \left[-{\rm ln}\left(\frac{f_{\rm min}}{\rm yr^{-1}}\right) \right]\cdot {\rm S/N}^2 \propto - {\rm ln}\left(\frac{\rm yr}{T_{\rm obs}}\right)\cdot {\rm S/N}^2
\end{aligned}
\end{cases}
\end{equation}
The angle between the horizontal axis and the principal axis of the posterior distribution is given by
\begin{equation}
\begin{aligned}
    \theta &= \frac{1}{2}\rm arctan \left(\frac{2\sigma_{A\gamma}}{\sigma_A^2 - \sigma_{\gamma}^2}\right) = \frac{1}{2}\rm arctan \left(\frac{2F^{-1}_{A\gamma}}{F^{-1}_{AA} - F^{-1}_{\gamma\gamma}}\right) \approx\\
    & \approx \frac{1}{2}\rm arctan \left(\frac{\left[- {\rm ln10}\cdot {\rm ln}\left(\frac{\rm yr}{T_{\rm obs}}\right)\cdot {\rm S/N}^2\right]^{-1}}{\left[(2{\rm ln10})^2 \cdot {\rm S/N}^2\right]^{-1} - \left[{\rm ln}\left(\frac{\rm yr}{T_{\rm obs}}\right)^2\cdot {\rm S/N}^2\right]^{-1}}\right) = \\
    & = \frac{1}{2}\rm arctan \left(\frac{-({\rm ln10})^{-1} \left[{\rm ln}\left(\frac{\rm yr}{T_{\rm obs}}\right)\right]^{-1}}{(2{\rm ln10})^{-2} - \left[{\rm ln}\left(\frac{\rm yr}{T_{\rm obs}}\right)\right]^{-2}}\right)
    \end{aligned}
\end{equation}

Since for $F^{-1}_{A\gamma}$ and $F^{-1}_{\gamma\gamma}$ we do not know the exact value, but only an approximation, we can write the exact expression for $\theta$ by adding two unknown variables
\begin{equation}\label{eq:theta}
    \theta = \frac{1}{2}\rm arctan \left(\frac{{\it a} \cdot \left[{\rm ln}\left(\frac{\rm yr}{T_{\rm obs}}\right)\right]^{-1}}{{\it b}- \left[{\rm ln}\left(\frac{\rm yr}{T_{\rm obs}}\right)\right]^{-2}}\right)
\end{equation}

\begin{figure}
   \centering
   \includegraphics[width=0.5\textwidth]{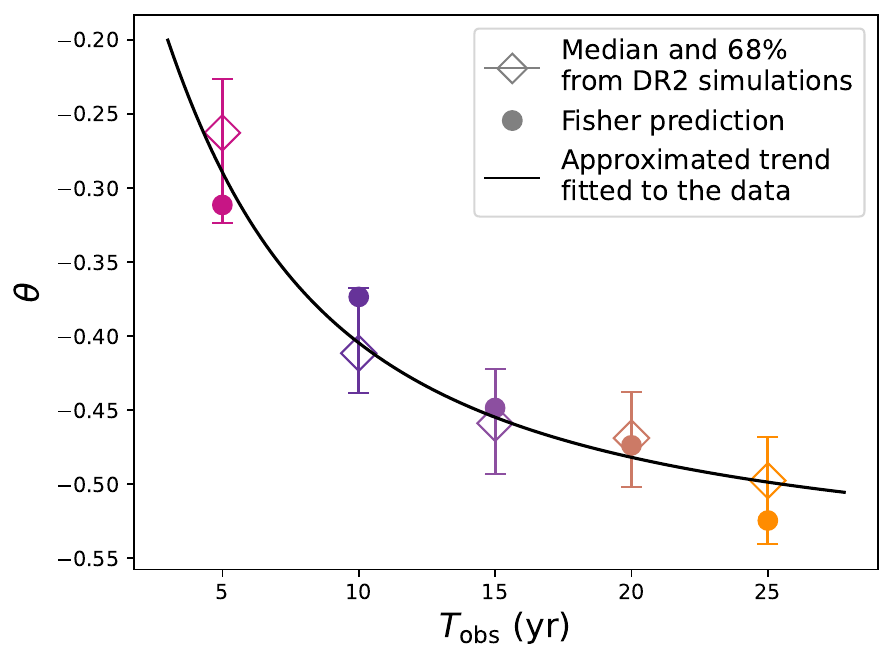}
   \caption{Orientation of the ($\gamma_{\rm GWB}$, $\log_{10}A_{\rm GWB}$) joint posterior distribution as a function of the observation time. This angle is computed from the 100 simulations of \texttt{DR2}, from the Fisher matrix formalism and with the approximated formula shown in \autoref{eq:theta} fitted to the data.}
   \label{fig:theta_anal}%
\end{figure}

By fitting this function to the $\theta$ measured from the 100 realisation of \texttt{DR2}, we find that the relation is in very good agreement with the behaviour of the data, as can be seen from \autoref{fig:theta_anal}.

\bibliographystyle{aa}
\bibliography{bibliography}

\end{document}